\documentclass[12pt]{iopart}
\usepackage{graphicx}
\usepackage{graphics}
\usepackage{amssymb}
\usepackage{amsmath}
\usepackage{cite}

\begin{document}

\title[Recurrence networks -- A novel paradigm for nonlinear time series analysis]{Recurrence networks -- A novel paradigm for nonlinear time series analysis}

\author{Reik V. Donner$^{1,2,3,4}$, Yong Zou$^3$, Jonathan F. Donges$^{3,5}$, Norbert Marwan$^3$, J\"urgen Kurths$^{3,5}$}

\address{$^1$Max Planck Institute for Physics of Complex Systems, Dresden, Germany \\
$^2$Institute for Transport and Economics, Dresden University of Technology, Germany \\
$^3$Potsdam Institute for Climate Impact Research, Potsdam, Germany \\
$^4$Graduate School of Science, Osaka Prefecture University, Sakai, Japan \\
$^5$Institute for Physics, Humboldt University of Berlin, Germany}

\ead{donner@vwi.tu-dresden.de}

\begin{abstract}
This paper presents a new approach for analysing structural properties of time series from complex systems. Starting from the concept of recurrences in phase space, the recurrence matrix of a time series is interpreted as the adjacency matrix of an associated complex network which links different points in time if the evolution of the considered states is very similar. A critical comparison of these recurrence networks with similar existing techniques is presented, revealing strong conceptual benefits of the new approach which can be considered as a unifying framework for transforming time series into complex networks that also includes other methods as special cases.

It is demonstrated that there are fundamental relationships between the topological properties of recurrence networks and the statistical properties of the phase space density of the underlying dynamical system. Hence, the network description yields new quantitative characteristics of the dynamical complexity of a time series, which substantially complement existing measures of recurrence quantification analysis.
\end{abstract}

\submitto{\NJP}
\maketitle

\section{Introduction} \label{sec1} 

Since the early stages of quantitative nonlinear sciences, numerous conceptual approaches have been introduced for studying the characteristic features of dynamical systems based on observational time series \cite{Abarbanel1996,Kantz1997,Diks1999,Sprott2003}. Popular methods that are increasingly used in a variety of applications (see, for example, \cite{Donner2008Book}) include (among others) Lyapunov exponents, fractal dimensions, symbolic discretisation, and measures of complexity such as entropies and quantities derived from them. All these techniques have in common that they quantify certain dynamically invariant phase space properties of the considered system based on temporally discretised realisations of individual trajectories.

As a particular concept the basic ideas of which originated in the pioneering work of Poincar\'e in the late 19th century \cite{Poincare1890}, the quantification of recurrence properties in phase space has recently attracted considerable interest~\cite{Saussol2002}. One particular reason for this is that these recurrences can be easily visualised (and subsequently quantified in a natural way) by means of so-called recurrence plots obtained from a single trajectory of the dynamical system under study \cite{Eckmann1987,Marwan2007}. When observing this trajectory as a scalar time series $x(t)$ ($t=1,\dots,N$), one may use a suitable $m$-dimensional time delay embedding of $x(t)$ with delay $\tau$ \cite{Takens1981}, $\mathbf{x}^{(m)}(t)=(x(t),x(t+\tau),\dots,x(t+(m-1)\tau))$, for obtaining a recurrence plot as a graphical representation of the binary recurrence matrix

\begin{equation}
R_{i,j}(\epsilon) = \Theta(\epsilon-\| \mathbf{x}_i-\mathbf{x}_j\|),
\end{equation}

\noindent where $\Theta(\cdot)$ is the Heaviside function, $\| \cdot\|$ denotes a suitable norm in the considered phase space, and $\epsilon$ is a threshold distance that should be reasonably smaller than the attractor diameter \cite{Marwan2007,Schinkel2008}. To simplify our notation, we have used the abbreviation $\mathbf{x}_i=\mathbf{x}^{(m)}(t=t_i)$ (with $t_i$ being the point in time associated with the $i$-th observation recorded in the time series\footnote{Note that unlike many other methods of time series analysis, the concept of recurrence plots does not require observations that are equally spaced in time.}) wherever appropriate.

Experimental time series often yield a recurrence plot displaying complex structures, in particular, with different properties of the non-interrupted diagonal and vertical structures (``lines''). A variety of statistical characteristics of the length distributions of these lines (such as maximum, mean, or Shannon entropy) can be used for defining additional quantitative measures that characterise different aspects of dynamic complexity of the studied system in more detail. This conceptual framework is known as recurrence quantification analysis (RQA)~\cite{Zbilut1992,Webber1994,Marwan2002} and is nowadays frequently applied to a variety of real-world
applications of time series analysis in various fields of research~\cite{Marwan2008}. However, most of these RQA measures are sensitive to the choice of embedding parameters, which are found to sometimes induce spurious correlations in a recurrence plot~\cite{Thiel2006}.

Recent studies have revealed that the fundamental invariant properties of a dynamical system (i.e., its correlation dimension $D_2$ and correlation entropy $K_2$) are conserved in the recurrence matrix~\cite{Thiel2004a}. Furthermore, it is found that the estimation of these invariants is independent of the particular embedding parameters. The recurrence plots preserve all the topologically relevant phase space information of the system, such that one can completely reconstruct a time series from its recurrence matrix (modulo some rescaling of its probability distribution function)~\cite{Thiel2004b,Robinson2007}.

A further appealing paradigm for analysing structural features of complex systems is based on their representation as complex networks of passive or active (i.e., mutually interacting) subsystems. For this purpose, classical graph theory has been systematically extended by a large variety of different statistical descriptors of the topological features of such networks on local, intermediate, and global scales \cite{Albert2002,Newman2003,daCosta2007}. These measures have been successfully applied for studying real-world networks in various scientific disciplines, including the structural properties of infrastructures \cite{Crucitti2004}, biological \cite{Zhou2006}, ecological \cite{Dunne2002}, and climate networks \cite{Donges2009epjst}, to give some prominent examples. The corresponding results have triggered substantial progress in our understanding of the interplay between structure and dynamics of such complex networks \cite{Wang2002,Boccaletti2006,Arenas2008}.

The great success of network theory in various fields of research has recently motivated first attempts to generalise this concept for a direct application to time series \cite{Zhang2006,Zhang2008,Xu2008,Shimada2008,Yang2008,Lacasa2008,Wu2008,Small2009,Donner2009EGU,Gao2009}. However, a substantial number of the recently suggested techniques have certain conceptual limitations, which make them suitable only for dealing with distinct types of problems. As an alternative that may provide a unifying conceptual and practical framework for nonlinear time series analysis using complex networks, we reconsider the concept of recurrences in phase space for defining complex network structures directly based on time series. For this purpose, it is straightforward to interpret the recurrence matrix $\mathbf{R}(\epsilon)$ as the adjacency matrix $\mathbf{A}(\epsilon)$ of an unweighted and undirected complex network, which we suggest to call the \textit{recurrence network} associated with a given time series. To be more specific, the associated adjacency matrix is given by 
\begin{equation}
A_{i,j} (\epsilon)= R_{i,j} (\epsilon) - \delta_{i,j},
\label{def:adjacency}
\end{equation}
\noindent 
where $\delta_{i,j}$ is the Kronecker delta introduced here in order to avoid artificial self-loops. A corresponding conceptual idea has recently been independently suggested by different authors \cite{Xu2008,Shimada2008,Wu2008,Small2009,Gao2009}, but not yet systematically studied. In this work, we however aim to give a rigorous derivation and detailed interpretation for a variety of quantitative characteristics of recurrence networks. It shall be noted that a generalisation to weighted networks (as partially studied in \cite{Zhang2006,Zhang2008}) is straightforward if the recurrence matrix is replaced by the associated distance matrix between pairs of states. In any case, recurrence networks referring to the mutual phase space distances of observational points on a single trajectory are spatial networks, i.e. fully embedded into an $m$-dimensional space, which has important implications for their specific topological features. We will raise this point in detail within this paper.

The consideration of recurrence plots as graphical representations of complex networks allows a reinterpretation of many network-theoretic measures in terms of characteristic phase space properties of a dynamical system. According to the ergodicity hypothesis, we suppose that one may gain full information about these properties by either ensembles of trajectories, or sufficiently long observations of a single trajectory. Following this line of ideas, we may approximate the (usually unknown) invariant density $p(\mathbf{x})$ (which is related with the associated invariant measure $\mu$ by $d\mu=p(\mathbf{x})d\mathbf{x}$) of the studied system by some empirical estimate $\hat{p}^{(\epsilon)}(\mathbf{x})$ obtained from a time series, where $\epsilon$ defines the level of coarse-graining of phase space involved in this procedure. Transforming the time series into a recurrence network then allows to quantitatively characterise the higher-order statistical properties of the invariant density $p(\mathbf{x})$ by means of complementary methods, i.e., network-theoretic measures.

According to the above argumentation, quantitative descriptors of the topological features of recurrence networks can be considered as novel measures within the framework of RQA. Our technique therefore exhibits additional deep insights into the phase space properties of dynamical systems directly related to their complex dynamics. Additionally, we will emphasise that we may also take seriously the duality of adjacency matrices of complex networks on the one hand, and recurrence matrices of dynamical systems on the other hand, which would also allow transferring concepts from dynamical systems theory (given that the corresponding recurrence plot based estimates are invariant under temporal reordering) to complex network theory. In this work, however, we will concentrate on a detailed discussion of how phase space properties can be further quantified in terms of network theory.

The remainder of this paper is organised as follows: Section \ref{sec:review} presents a critical review of existing approaches for extracting complex networks from time series, including a comprehensive discussion of their potentials and potential problems (with a special emphasis on how to interpret the resulting networks' topological properties). The concept of recurrence networks as a natural alternative is further discussed in Section \ref{sec:recnet}. In particular, it is demonstrated that many network-theoretic measures yield sophisticated quantitative characteristics corresponding to certain phase space properties of a dynamical system that have not yet been explicitly studied in terms of other dynamical invariants or measures of complexity based on RQA. 
In order to support our theoretical considerations, Section \ref{sec:examples} provides some examples of how different network-theoretic measures reveal certain phase space properties of various dynamical systems. Finally, we summarise our main results and outline some future directions of further methodological developments based on our proposed technique.

\section{Approaches for transforming time series into complex networks - A comparative review} \label{sec:review}

In this section, a review and classification of existing approaches for studying the properties of time series by means of complex networks approaches is presented (see Tab.~\ref{tab:methods}). In particular, strengths and possible limitations of the existing approaches will be briefly discussed.

\begin{table}[t]
\small \begin{tabular}{lcr}
Method & Vertex & Edge \\
\hline
Coarse-graining (2005) \cite{Gao2005,PLi2006} & Discrete state $s(t)$ & Equality of states \\
Cycle networks (2006) \cite{Zhang2006} & Cycle & Correlation between cycles \\
Correlation method (2008) \cite{Yang2008} & State $\textbf{x}(t)$ & Correlation between state vectors \\
Visibility graph (2008) \cite{Lacasa2008} & Scalar state $x(t)$ & Mutual visibility of states \\
Neighbourhood network (2008) \cite{Xu2008,Shimada2008} & State $\textbf{x}(t)$  & Recurrence of states (Mass) \\
Recurrence network (2008) \cite{Wu2008,Gao2009} & State $\textbf{x}(t)$ & Recurrence of states (Volume) \\
\end{tabular}
\normalsize
\caption{\label{VertexEdgeTable}Summary of the definitions of vertices and the criteria for the existence of edges in existing complex network approaches to time series analysis (given in chronological order).}
\label{tab:methods}
\end{table}

\subsection{Coarse-graining of phase space}

The simplest possible method for transforming a time series into a complex network representation is coarse-graining its range into a suitable set of classes and considering the transition probabilities between these classes in terms of a weighted network. In general, the underlying concept of symbolic dynamics \cite{Robinson1999} allows characterising the properties of dynamical systems based on a partitioning of its phase space, yielding a transformation of every possible trajectory into an inifinite sequence of abstract symbols. Formally, the application of resulting methods of symbolic time series analysis (such as mutual information or entropic quantities) requires the existence of a generating partition which corresponds to a \textit{unique} assignment of symbolic sequences (i.e., sequences of class identifiers) to every trajectory of the system. Note that this prerequisite is usually violated in real-world applications due to the presence of noise, however, even in the ideal noise-free case generating partitions do either not exist or can hardly be estimated (see \cite{Hirata2004} and references therein). Nevertheless, applications of symbolic time series analysis have recently attracted considerable interest in numerous applications \cite{Daw2003,Donner2008EPJST}.

Gao \textit{et~al.} \cite{Gao2005,Li2006,Li2007,Gao2007} used a specific coarse-graining of the phase space for studying transitions between traffic states in different cellular automaton models. They however restricted their attention to the consideration of degree distributions. Recently, their idea was generalised to weighted networks by Zheng and Gao \cite{Zheng2008}. In a similar way, Li \textit{et al.} \cite{PLi2006,PLi2007} used a corresponding approach for a coarse-grained analysis of stock exchange time series. In particular, their research mainly focussed on the identification of vertices with the highest relevance for information transfer, which have been quantified in terms of betweenness centrality and inverse participation ratio of the individual network vertices.

The main disadvantage of the coarse-graining approach is that it may yield a significant loss of information on small amplitude variations. In particular, two observations with even very similar values may not be considered to belong to the same class if they are just separated by a class boundary. This might influence the quantitative features of a corresponding network, as it is not exclusively determined by the widths of the individual classes, but also their specific definition. In this respect, the recurrence network approach introduced in this work is more objective as it only depends on a single parameter $\epsilon$. Note, however, that coarse-graining might be a valid approach in case of noisy real-world time series, where extraction of dynamically relevant information hidden by the noise can be supported by grouping the data.

\subsection{Cycle networks}

In 2006, Zhang and Small \cite{Zhang2006,Zhang2008,Xu2008} suggested to study the topological features of pseudo-periodic time series (representing, for example, the dynamics of chaotic oscillators like the Lorenz or R\"ossler systems) by means of complex networks. For this purpose, individual cycles (defined by minima or maxima of the studied time series) have been considered as vertices of a network, and the connectivity of pairs of vertices has been established by considering a generalisation of the correlation coefficient to cycles of possibly different length or, alternatively, their phase space distance. 

A potential point of criticism to this method is that the definition of a cycle is not necessarily straightforward in complex oscillatory systems. In \cite{Zhang2006,Zhang2008,Xu2008}, the authors mainly considered nonlinear oscillators in their phase-coherent regimes, however, it is not clear how a cycle could be defined for non-phase-coherent oscillations, for example, in the Funnel regime of the R\"ossler system. The same problem arises for systems with multiple time scales which are hence hard to treat this way. Furthermore, it is not intuitive how one can interpret correlations of cycles, since the values of the corresponding measures are not exclusively determined by the proximity of the corresponding parts of the trajectory in phase space, but depend also on the ``lengths'' of the cycles (in terms of the number of states) that may vary due to the discrete sampling. Hence, one may find rather low correlations between two cycles although the two parts of the trajectory are actually close to each other.

\subsection{Correlation networks of embedded state vectors}\label{sec:yang}

A generalisation of the technique used by Zhang and Small that can also be applied to time series without obvious oscillatory components has been suggested by Yang and Yang \cite{Yang2008} using a simple embedding of an arbitrary time series. In their formalism, individual state vectors $\mathbf{x}^{(m)}_i$ in the $m$-dimensional phase space of the embedded variables are considered as vertices, from which a Pearson correlation coefficient $r_{i,j} = r(\mathbf{x}^{(m)}_i,\mathbf{x}^{(m)}_j)$ can be easily computed. If $r_{i,j}$ is larger than a given threshold, vertices $i$ and $j$ are considered to be connected. The approach of Yang and Yang has recently been reinvented by Gao and Jin \cite{Gao2009,Gao2009b} in terms of so-called fluid-dynamic complex networks (FDCN) that have been successfully applied for characterising the nonlinear dynamics of conductance fluctuating signals in a gas-liquid two-phase flow.

One potential conceptual problem of this particular technique is that the consideration of correlation coefficients between two phase space vectors usually requires a sufficiently large embedding dimension $m$ for a proper estimation with low uncertainty (more specifically, the standard error of the correlation coefficient is approximately proportional to $1/\sqrt{m-1}$). Hence, local information about the short-term dynamics captured in a time series might get lost when following this approach. Even more, since embedding is known to induce spurious correlations to a system under study, the results of the correlation method of network construction may suffer from related effects.

With respect to the interpretation of the resulting network patterns, one has to note that for vertices corresponding to mutually overlapping time series segments, the consideration of correlation coefficients, as applied in both papers cited above, corresponds to studying the local auto-correlation function of the signal. Hence, the presence of edges between these vertices is exclusively determined by linear correlations within the signals. In principle, we might think of replacing the correlation coefficient by other measures of interrelationships such as the mutual information, that are also sensitive to general statistical dependences \cite{Donges2009epjst,Donges2009epl,Donner2010}; however, the appropriate estimation of such nonlinear quantities would require an even considerably larger amount of data, i.e., a very large embedding dimension $m$.

Finally, when studying time series with pronounced cycles (like trajectories of the R\"ossler or Lorenz systems), there may then be different cycles included in one embedding vector (depending on the sampling rate), which casts additional doubts with respect to the direct interpretability of the resulting network properties.

\subsection{Visibility graphs}

An alternative to the latter two threshold-based concepts has been suggested by Lacasa \textit{et~al.} \cite{Lacasa2008} in terms of the so-called visibility graph. In this formalism, individual observations are considered as vertices, and edges are introduced whenever a partial convexity constraint is fulfilled, i.e.~$x(t_a)$ and $x(t_b)$ are connected if for all states $x(t_c)$ with $t_a<t_c<t_b$, 

\begin{equation}
\frac{x_c-x_a}{t_a-t_c}>\frac{x_a-x_b}{t_b-t_a}
\end{equation}

\noindent holds. Visibility graphs have been used to study the behaviour of certain fractal as well as multifractal stochastic processes \cite{Ni2008,Lacasa2009}, energy dissipation in three-dimensional turbulence \cite{Liu2009}, and the nonlinear properties of exchange rate time series \cite{Yang2009}.

The network corresponding to a visibility graph is easily established and allows to distinguish between different types of systems. However, there is no straightforward interpretation of the convexity constraint in terms of phase space properties of the considered system. Moreover, the application of this approach is restricted to univariate time series, while at least the approach by Yang and Yang could in principle be easily generalised to multivariate time series.

\subsection{Complex networks based on neighbourhood relations in phase space}

As it has already been mentioned, the transformation of time series in terms of neighbourhood relationships has already been discussed by different authors. In particular, there are two possible approaches, that can directly be related to slightly different definitions of recurrence plots \cite{Marwan2007}:

On the one hand, a neighbourhood can be defined by a fixed number of nearest neighbours of a single observation, i.e., a constant ``mass'' of the considered environments \cite{Xu2008,Shimada2008}. We refer to this method as a \textit{neighbourhood network} in phase space. This setting implies that the degrees $k_v$ of all vertices in the network are kept fixed at the same value. Hence, information about the local geometry of the phase space, which is mainly determined by the invariant density $p(\mathbf{x})$, cannot be directly obtained by most traditional complex network measures (see Sec.~\ref{sec:recnet})\footnote{As an alternative measure, one could consider the maximum distance of the $k$-th nearest neighbour as a measure for phase space density.}. Note that the adjacency matrix of a neighbourhood network is in general not symmetric, i.e., the fact that a vertex $j$ is among the $k$ nearest neighbours of a vertex $i$ does \textit{not} imply that $i$ is also among the $k$ nearest neighbours of $j$. Hence, neighbourhood networks can be formally considered as (partially) directed networks.

On the other hand, one may define the neighbourhood of a single point in phase space (represented by a certain observation) by a fixed phase space distance, i.e., considering a constant ``volume'' \cite{Wu2008,Gao2009}. This approach has the advantage that the degree centrality $k_v$ gives direct information about the local phase space density (see Section \ref{sec:recnet}). Gao and Jin \cite{Gao2009} termed a corresponding approach as fluid-structure complex networks (FSCN) and used it for analysing gas-liquid two-phase flow and the Lorenz system as a toy model in terms of link density. In addition, they related their results to the presence of unstable periodic orbits in a dynamical system. We will come back to this point in Section \ref{sec:examples_spatial}.

The consideration of neighbourhood relationships within a fixed phase space volume corresponds to the standard definition of a recurrence plot as mentioned in the introductory section. Hence, the resulting networks will be referred to as recurrence networks in the following. In particular, all arguments provided in the remainder of this paper for recurrence networks are based on the idea of a fixed volume of the considered neighbourhoods rather than a fixed mass and may not be directly generalised to the other case.

\section{Quantitative assessment of recurrence networks} \label{sec:recnet}

\begin{table}
\begin{tabular}{lr}
Recurrence network & Phase space\\
\hline
Vertex & State $\textbf{x}(t)$ \\
Edge & Recurrence of states \\
Path & Overlapping sequence of $\epsilon$-balls\\
\end{tabular}
\caption{\label{GeometricalObjectsTable}Relationships between recurrence network entities and corresponding geometrical objects and their properties in phase space.}
\end{table}

Many of the already existing methods for transforming time series into complex networks that have been discussed in Section \ref{sec:review} suffer (among other problems) from the fact that there is no direct link between the local properties of the considered time series and the topology of the resulting complex networks. In particular, the concepts used for defining both vertices and edges of the networks, which differ across the various techniques, are in some cases rather artificial from a dynamical systems point of view (Table \ref{VertexEdgeTable}).

In contrast to the other recently suggested approaches, the identification of a recurrence matrix with the adjacency matrix of a complex network is a straightforward and natural idea that conserves many local properties of the considered time series. In particular, individual values of the respective observable can be directly considered as vertices of the recurrence network (similar to the visibility graph concept), while the existence of an edge serves as an indicator of a recurrence, i.e., pairs of states whose values do not differ by more than a small value $\epsilon$ in terms of a suitable norm in phase space. 

It should be noted that the recurrence networks approach followed in this work is not the only concept that combines basic ideas of recurrence plots and complex networks. Apart from the neighbourhood networks originated in the idea of a fixed recurrence rate (i.e., a fixed mass of the considered neighbourhoods), the idea of considering a threshold value to the proximity of two vertices can also be found in other previously suggested methods. In particular, the coarse-graining approach is equivalent to considering recurrence plots of discrete-valued observables with a threshold of $\epsilon=0$. Moreover, the correlation method of Yang and Yang \cite{Yang2008} (see Sec.~\ref{sec:yang}) can also be considered as being based on a recurrence plot where the usual metric distance has been replaced by the correlation distance \cite{Donner2008Bookchapter}
\begin{equation}
d_C(\mathbf{x}_i,\mathbf{x}_j)=1-r_{i,j}.
\end{equation}
\noindent 
Note, however, that the advantage of considering the concept of recurrences defined in terms of metric distances in phase space instead of correlations is that it allows for creating networks based even on individual states without any embedding or consideration of groups of states. On the one hand, this independence from a particular embedding may be beneficial when dynamical invariants of the studied system are of interest. On the other hand, the statistical properties of the resulting recurrence networks reflect \textit{exclusively} the invariant density of states in phase space (in terms of certain higher-order statistics), because time-ordering information is lost in this framework. Hence, it is hardly possible to distinguish between, e.g., deterministic and stochastic dynamics. Here, additional embedding might in fact provide a feasible solution to the corresponding identification problem. 

Following the above considerations, it can be argued that the concept of recurrence networks yields a general framework for transferring time series into complex networks in a dynamically meaningful way. In particular, this approach can be applied (i) to both univariate as well as multivariate time series (phase space trajectories) (ii) with and without pronounced oscillatory components and (iii) with as well as without embedding. Consequently, unlike for most existing techniques, there are no fundamental restrictions with respect to its practical applicability to arbitrary time series.

While the definitions of edges and vertices in our approach have already been given above (Table \ref{VertexEdgeTable}), we now provide a geometrical interpretation of a third important network entity, the path, within the framework of recurrence networks (Table \ref{GeometricalObjectsTable}). A path between two vertices $i$ to $j$ in a simple graph without multiple edges can be written as an ordered sequence of the vertices it contains, i.e., $(i,k_1,\dots,k_{l_{i,j}-1},j)$, where the associated number of edges $l_{i,j}$ measures the length of the path. In phase space, a path in the recurrence network is hence defined as a sequence of mutually overlapping $\epsilon$-balls $B_\epsilon(\mathbf{x}_i),B_\epsilon(\mathbf{x}_{k_1}),\dots,B_\epsilon(\mathbf{x}_{k_{l_{i,j}-1}}),B_\epsilon(\mathbf{x}_j)$, where $B_\epsilon(\mathbf{x}_i) \cap B_\epsilon(\mathbf{x}t_{k_1}) \neq \emptyset, \dots, B_\epsilon(\mathbf{x}_{k_{l_{i,j}-1}}) \cap B_\epsilon(\mathbf{x}_j) \neq \emptyset$ \footnote{An $\epsilon$-ball centered at state vector $\mathbf{x}$ is defined as the open set $B_\epsilon(\mathbf{x}) = \{\mathbf{y} \in \mathbb{R}^m:||\mathbf{x}-\mathbf{y}|| < \epsilon\}$.}. 

Due to the natural interpretation of vertices, edges and paths, the topological characteristics of a recurrence network closely capture the fundamental phase space properties of the dynamical system that has generated the considered time series. In the following, we will present a detailed analysis of the corresponding analogies for different network properties that are defined on a local (i.e.~considering only the direct neighbourhood of a vertex), intermediate (i.e.~considering the neighbourhood of the neighbours of a vertex), and global (i.e.~considering all vertices) scale (Table \ref{NetworkRQATable})\footnote{Alternatively, one may classify the corresponding phase space properties according to the fact whether they refer to individual points, small regions, or the entire phase space. In this respect, measures related to a single vertex $v$ (centralities, local clustering coefficient, local degree anomaly) give local, those related to a specific edge $(i,j)$ (shortest path length, matching index, edge betweenness) intermediate, and all others global information about the phase space properties.}. It has to be emphasised that these quantities can be considered as (partly novel and complementary) measures in the framework of RQA.

\begin{table}
\begin{tabular}{llr}
Scale & Recurrence network & Phase space\\
\hline
Local & Edge density $\rho$ & Global recurrence rate $RR$\\
& Degree centrality $k_v$ & Local recurrence rate $RR_v$\\
\hline
Intermediate & Clustering coefficient $\mathcal{C}$ & Invariant objects\\
& Local degree anomaly $\Delta k_v$ & Local heterogeneity of phase space density\\
& Assortativity $\mathcal{R}$ & Continuity of phase space density\\
& Matching index $\mu_{i,j}$ & Twinness of $i,j$\\
\hline
Global & Average path length $\mathcal{L}$ & Mean phase space separation $\left<d_{i,j}\right>_{i,j}$\\
& Network diameter $\mathcal{D}$ & Phase space diameter $\Delta$\\
& Closeness centrality $c_v$ & Local centeredness in phase space \\
& Betweenness centrality $b_v$ & Local attractor fractionation\\
\end{tabular}
\caption{\label{NetworkRQATable}Correspondence between recurrence network measures and phase space properties. Specific terms are discussed in the text.}
\end{table}

\subsection{Local network properties}

\subsubsection{Degree centrality (local recurrence rate).}

As a first measure that allows to quantify the importance of a vertex in a complex network, the degree centrality of a vertex $v$, $k_v$, is defined as the number of neighbours, i.e.~the number of vertices $i\neq v$ that are directly connected with $v$:
\begin{equation} \label{degree}
  k_v = \sum_{i=1}^{N} A_{v,i}.
\end{equation}
\noindent
Note that in general, the sum is taken over all $i\neq v$. However, according to our definition (\ref{def:adjacency}), we skip the corresponding condition in the following. Normalising this measure by the maximum number of possible connections, $N-1$, one gets the local connectivity
\begin{equation} \label{recrate}
\rho_v=\frac{1}{N-1} \sum_{i=1}^{N} A_{v,i}=RR_v,
\end{equation}
\noindent
which, from the recurrence plot point of view, corresponds to the local recurrence rate $RR_v$ of the state $v$. Thus, the degree centrality and local connectivity yield an estimator for the local phase space density, since for a vertex $v$ located at position $\mathbf{x}_v$ in phase space,
\begin{equation}
\frac{1}{N}(k_v(\epsilon)+1) \approx \int_{B_{\epsilon}(\mathbf{x}_v)} d\mathbf{x}\ p(\mathbf{x}) \approx (2\epsilon)^m p(\mathbf{x}_v)
\end{equation}
\noindent
(when using the maximum norm) and, hence,
\begin{equation}
\hat{p}(\mathbf{x}_v)=\lim_{\epsilon\to 0} \lim_{N\to\infty} \frac{k_v+1}{(2\epsilon)^m N}.
\end{equation}

In complex network studies, one is often interested in the frequency distribution of degree centralities, $P(k)$, in particular, the presence of an algebraic scaling behaviour, which is characteristic for scale-free networks \cite{Albert2002}. However, although several authors have recently focussed their attention on this characteristic obtained from different types of complex networks derived from time series \cite{Zhang2006,Yang2008,Lacasa2008,Wu2008,Gao2009,Gao2005,Li2006,Li2007,Gao2007,Ni2008,Lacasa2009,Liu2009}, we would like to underline that for a complete characterisation of the phase space properties of a dynamical system, one should prefer studying not only degree centralities, but also other higher-order statistical measures.

\subsubsection{Edge density (global recurrence rate).}

In some situations, it is useful not to consider the full distribution of degree centralities in a network, but to focus on the mean degree of all vertices
\begin{equation}
\left<k\right>=\frac{1}{N}\sum_{v=1}^N k_v=\frac{2L}{N},
\end{equation}
\noindent
as a simple characteristic quantity of this distribution, where
\begin{equation}
L=\sum_{i<j} A_{i,j}=\rho \frac{N(N-1)}{2}
\end{equation}
\noindent
is the total number of edges in the recurrence network. The mean degree centrality $\left<k\right>$ is directly proportional to the edge density $\rho$ of the network or, alternatively, its recurrence plot equivalent, the global recurrence rate $RR$,
\begin{equation}
\label{rho_density}
\begin{split}
  \rho(\epsilon) & = \frac{1}{N} \sum_{v=1}^{N} \rho_v(\epsilon) =  
  \frac{2}{N(N-1)} \sum_{v<i} A_{v,i}(\epsilon) \\
       & = \frac{2}{N(N-1)} \sum_{v<i} \Theta(\epsilon-\|\mathbf{x}_{v} - \mathbf{x}_{i} \|) \\
       & = RR(\epsilon) = C_2(\epsilon) \sim {\epsilon}^{-D_2}.
\end{split}
\end{equation}
Note that the recurrence rate coincides with the definition of the correlation integral $C_2(\epsilon)$, which is commonly used to estimate the correlation dimension $D_2$, for example, using the Grassberger-Procaccia algorithm \cite{Grassberger1983}. 

The connection between the edge density and the correlation dimension can be understood by the fact that the local recurrence rate $RR_v$ of a vertex $v$ corresponds to the measure of a $m$-dimensional ball $B_\epsilon(\mathbf{x}_v)$ of radius $\epsilon$ centered at the point $\mathbf{x}_v$ in the $m$-dimensional phase space in the limit that time goes to infinity ($N \to \infty$). When considering the Euclidean norm as a distance measure in phase space, these balls are defined as hyperspheres, for the maximum norm as hypercubes etc. Then, the pointwise (information) dimension of the probability measure $\mu$ at $\mathbf{x}_{v}$ is defined as 
$D_p(\mathbf{x}_{v}) = - \lim_{\epsilon \to 0} \left(\ln \mu(B_\epsilon(\mathbf{x}_v))/\ln \epsilon\right)$ \cite{Sprott2003}. Due to the heterogeneity of the phase space visited by the trajectory (i.e., the non-uniform phase space density that results in different degree centralities $k_v$ in different parts of this space), the proper estimation of $D_p$ is a nontrivial task and often requires expensive computational power and a high data quality and quantity. Thus, one may expect a better statistics for $D_2$, since it more heavily weights regions of the phase space which have a higher probability measure $\mu$. Though the correlation integral has been well established in the literature for estimating the correlation dimension, we point out the improvement in estimating $D_2$ based on the diagonal lines in $R_{i,j}(\epsilon)$, which yields an algorithm that is independent of the embedding parameters~\cite{Thiel2004a}. Consequently, the recurrence network representation $A_{i,j}$ of a time series fully conserves the geometric properties of the phase space of the underlying dynamical system.

\subsection{Intermediate scale network properties}

\subsubsection{Local clustering coefficient.} 

The clustering coefficient of a vertex $v$, $\mathcal{C}_v$, characterises the density of connections in the direct neighbourhood of this vertex in terms of the density of connections between all vertices that are incident with $v$. In many networks, such loop structures formed by three vertices occur more often than one would expect for a completely random network. Hence, high clustering coefficients reveal a specific type of structure in a network, which is related to the cliquishness of a vertex \cite{daCosta2007}.

In this work, we consider the definition of clustering coefficient proposed by Watts and Strogatz \cite{daCosta2007}, 
\begin{equation}
\mathcal{C}_v=\frac{2}{k_v(k_v-1)} N^{\Delta}_v, \label{cc_local}
\end{equation}
\noindent
where $N^{\Delta}_v$ is the total number of closed triangles including vertex $v$, which is bound by the maximum possible value of $k_v(k_v-1)/2$. For vertices of degree $k_v=0$ or $1$ (isolated or tree-like points, respectively), the clustering coefficient is defined as $\mathcal{C}_v=0$, as such vertices cannot participate in triangles by definition.

Equation (\ref{cc_local}) can be rewritten in terms of conditional probabilities as
\begin{equation} \label{cc_ws}
\mathcal{C}_v=P(A_{i,j}=1|A_{v,i}=1,A_{v,j}=1)=\frac{P(A_{i,j}=1,A_{v,i}=1,A_{v,j}=1)}{P(A_{v,i}=1,A_{v,j}=1)}
\end{equation}
\noindent
using Bayes' theorem, with
\begin{equation}
P(A_{v,i}=1,A_{v,j}=1)=\frac{1}{(N-1)(N-2)}\sum_{i=1}^N \sum_{j=1,j\neq i}^N A_{v,i} A_{v,j}
\end{equation}
\noindent
and a similar expression for $P(A_{i,j}=1,A_{v,i}=1,A_{v,j}=1)$. As for a recurrence network, the value of $A_{i,j}$ depends only on the phase space distance and the choice of $\epsilon$, the latter relationship may be used to derive analytical results at least for one-dimensional systems based on their invariant density. Corresponding details can be found in \ref{app_1dmap}, including the corresponding treatment of the Bernoulli and logistic maps as specific examples.

\subsubsection{Global clustering coefficient.} 

As for the degree centrality, one may consider the average value of the clustering coefficient taken over all vertices of a network, the so-called global clustering coefficient
\begin{equation}
\mathcal{C}=\frac{1}{N} \sum_{v=1}^N \mathcal{C}_v,
\end{equation}
\noindent
as a global characteristic parameter of the topology of a network. One expects that the value of $\mathcal{C}$ is -- for a given dynamical system with a phase space density $p(\mathbf{x})$ -- in the asymptotic limit $N\to\infty$ exclusively determined by the choice of $\epsilon$, which determines the scale of resolution. A more detailed discussion of the corresponding effects and their implications for certain model systems will be given in Section \ref{sec:examples}.

\subsubsection{Mean nearest neighbour degree.}

The mean nearest neighbour degree $k^{nn}_v$ of vertex $v$ gives the average degree in the neighbourhood of $v$,
\begin{equation}
k^{nn}_v = \frac{1}{k_v} \sum_{i=1}^N A_{v,i} k_i.
\end{equation}
\noindent 
The degree centrality $k_v$ is a measure of the density of states in the immediate neighbourhood of state $v$, whereas $k^{nn}_v$ can be interpreted to indicate the mean density of states in the next neighbourhood (next topological shell of neighbours) of state $v$. Hence, both measures taken together contain information about the local density anomaly in the vicinity of $v$, which we propose to measure by the local degree anomaly
\begin{equation}
\Delta k_v = k_v - k^{nn}_v.
\end{equation}
\noindent
Vertices with a positive degree anomaly ($\Delta k_v>0$) hence indicate local maxima of phase space density, while such which $\Delta k_v<0$ correspond to local density minima. Hence, the local degree anomaly may be considered as a proxy for the local heterogeneity of the phase space density. In a similar way, the average absolute value of the local degree anomaly, $\left<|\Delta k_v|\right>_v$, serves as a measure for the overall spatial heterogeneity of the phase space density profile.

\subsubsection{Assortativity.} 

A network is called assortative if vertices tend to connect preferentially to vertices of a similar degree $k$. On the other hand, it is called disassortative if vertices of high degree prefer to connect to vertices of low degree, and vice versa. Hence, assortativity can be quantified by the Pearson correlation coefficient of the vertex degrees on both ends of all edges \cite{daCosta2007,Newman2002},
\begin{equation}
\mathcal{R} = \frac{\frac{1}{L} \sum_{j>i} k_i k_j A_{i,j} - \left[ \frac{1}{L} \sum_{j>i} \frac{1}{2} (k_i+k_j)A_{i,j} \right]^2}{\frac{1}{L} \sum_{j>i} \frac{1}{2} (k_i^2+k_j^2)A_{i,j} - \left[ \frac{1}{L} \sum_{j>i} \frac{1}{2} (k_i+k_j)A_{i,j} \right]^2}.
\end{equation}
\noindent 
If the density of states in phase space hardly varies within an $\epsilon$-ball, the degrees on either ends of an edge will tend to be similar and hence the assortativity coefficient $\mathcal{R}$ will be positive. This means, that the more continuous and slowly changing the density of states is, the closer $\mathcal{R}$ will be to its maximum value one. Within the framework of recurrence networks, $\mathcal{R}$ can hence be interpreted as a measure of the continuity of the density of states or put differently, of the fragmentation of the attractor. Note that this aspect has not yet been specifically addressed by other nonlinear measures, in particular, within the RQA framework.

\subsubsection{Matching index (twinness).} \label{sec:mi}

The overlap of the neighbourhood spaces of two vertices $i,j$ is measured by the matching index

\begin{equation}
\mu_{i,j} = \frac{\sum_{l=1}^N A_{i,l} A_{j,l}}{k_i + k_j - \sum_{l=1}^N A_{i,l} A_{j,l}},
\end{equation}

\noindent where $\mu_{i,j}=0$ if there are no common neighbours, and $\mu_{i,j}=1$ if the neighbourhoods coincide \cite{daCosta2007}. Using the notion of $\epsilon$-balls around points in phase space, one may alternatively write
\begin{equation}
\mu_{i,j}\propto \frac{\mu(B_{\epsilon}(\mathbf{x}_i)\cap B_{\epsilon}(\mathbf{x}_j))}{\mu(B_{\epsilon}(\mathbf{x}_i))+\mu(B_{\epsilon}(\mathbf{x}_i))-\mu(B_{\epsilon}(\mathbf{x}_i)\cap B_{\epsilon}(\mathbf{x}_j))}.
\end{equation}

Due to the spatial constraints of the recurrence network, the neighbourhood spaces of $i,j$ can only overlap if 
\begin{equation}
d_{i,j}=\|\mathbf{x}_i-\mathbf{x}_j\| \leq 2 \epsilon,
\end{equation}
\noindent
i.e., $\mu_{i,j}=0$ for all pairs of vertices $(i,j)$ with $d_{i,j} > 2 \epsilon$. Moreover, the matching index $\mu_{i,j}$ decreases on average with an increasing spatial distance $d_{i,j}$ between the two considered states. Note that since $A_{i,j}=0$ already for $d_{i,j}>\epsilon$, there may be unconnected points with a matching index $\mu_{i,j}>0$.

The matching index of pairs of vertices in a recurrence network is closely related to the concept of \textit{twins} \cite{Thiel2006b}, which has recently been successfully applied for constructing surrogate data (twin surrogates) in the context of statistical hypothesis testing for the presence of complex synchronisation \cite{Romano2009,vanLeeuwen2009}. Twins are defined as two states of a complex system that share the same neighbourhood in phase space, i.e., the two vertices of the recurrence network representing these states have a matching index $\mu_{i,j}=1$. Hence, the matching index can be used for identifying candidates for twins. Note that pairs of vertices $i$ and $j$ in a recurrence network with $\mu_{i,j}\lesssim 1$ can still be considered as \textit{potential} twins, since $\mu_{i,j}=1$ may in some cases be approached by only slight changes of the threshold $\epsilon$. Consequently, we suggest interpreting the matching index as a measure of the \textit{twinness} of $i$ and $j$. Furthermore, it should be noted that adjacent pairs of edges $(i,j)$ ($A_{i,j}=1$) with a low matching index $\mu_{i,j}\simeq 0$ connect two distinct regions of the attractor and may therefore be indicative of geometrical bottlenecks in the dynamics (cf. our discussion of the betweenness centrality in Section \ref{sec:bc}).

\subsection{Global network properties}

\subsubsection{Shortest path length.}

As we consider recurrence networks as undirected and unweighted, we assume all edges to be of unit length in terms of graph (geodesic) distance. Consequently, the distance between any two vertices of the network is defined as the length of the shortest path between them. Note that time information is lost after transforming the trajectory into a network presentation. Therefore, the terminology of the shortest path length $l_{i,j}$ in the recurrence network reflects the minimum number of edges that have to be passed on a graph between a vertex $i$ to a vertex $j$. In the same spirit, $l_{i,j}$ is related to the distance of states $i$ and $j$ in phase space. 
\begin{figure}[htbp]
  \centering
  \includegraphics[width=0.75\columnwidth]{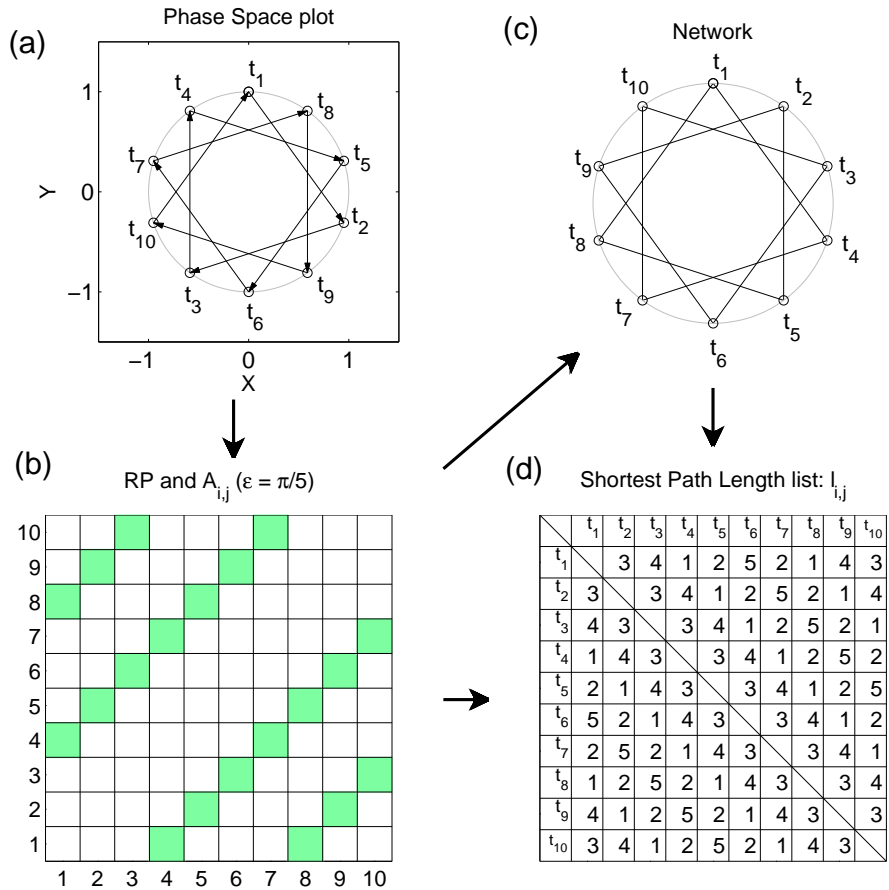}
\caption{\small Schematic representation of the transformation of (a) a periodic trajectory in phase space (arrows indicate the temporal order of observations) into (c) the recurrence network (lines illustrate the mutual neighbourhood relations). (b) gives the associated recurrence plot representation. Table (d) lists the resulting shortest path lengths between any two vertices.\label{periodic_list} }
\end{figure}
\begin{figure}[htbp]
  \centering
  \includegraphics[width=0.75\columnwidth]{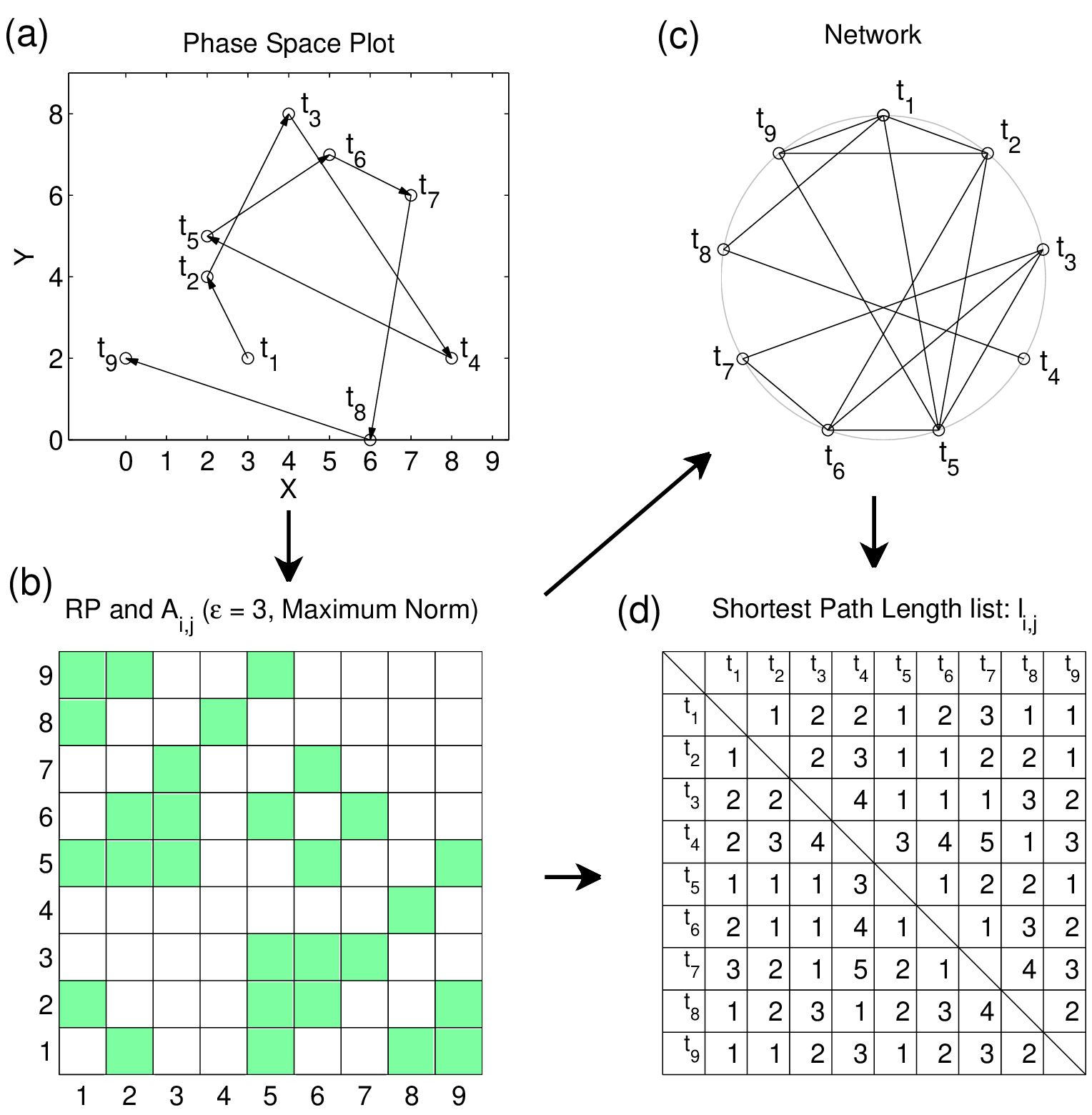}
\caption{\small Same as in Fig.~\ref{periodic_list} for a general non-periodic trajectory $(3,2,4,8,2,5,7,6,0,2)$ embedded in a two-dimensional phase space $(X,Y)=(x_t,x_{t+1})$. \label{nonperiodic_list}}
\end{figure}

In order to better understand the meaning of shortest path lengths, let us study their calculation for two toy model series: First, we consider a periodic trajectory $x(t) = \sin (6\pi\cdot 0.1t)$,
$y(t) = \cos (6\pi\cdot 0.1t)$, with $t=0,1,\cdots,10$, i.e.~there are $N=10$ points in the phase space
(Fig.~\ref{periodic_list}(a)). The corresponding recurrence plot is shown for $\epsilon=\frac{\pi}{5}$ (Fig.~\ref{periodic_list}(b)). As it has already been mentioned above, the recurrence matrix $R_{i,j}$ and the adjacency matrix $A_{i,j}$ of the associated recurrence network are basically equivalent. Adopting a common visualisation of connectivity patterns from the literature on complex networks, we illustrate the recurrences of the considered model time series by placing the individual observations (vertices) on a circle 
with equal common distances (Fig.~\ref{periodic_list}(c)). In this representation, the shortest path length (in the network sense) between two vertices $i$ and $j$ corresponds to the smallest number of ``jumps'' in phase space via pairs of neighbours (i.e. recurrences) in phase space. Obviously, the number of such jumps is determined by the prescribed value of $\epsilon$ and the spatial distance between $i$ and $j$. For instance, the shortest path from vertex $i=1$ to $j=10$ is $l_{1,10} = 3$ as indicated by the matrix of mutual shortest path lengths (Fig.~\ref{periodic_list}(d)). Note that this list is symmetric by definition, i.e. $l_{i,j} = l_{j,i}$). The same heuristic analysis can also be performed for a general nonperiodic trajectory in phase space as shown in Fig.~\ref{nonperiodic_list}(a-d). 

We wish to underline that the terminology of shortest path lengths in networks does not have a direct relevance to the dynamical evolution of the observed system. In contrast, $l_{i,j}$ measures distances in phase space (among a discrete set of points on the attractor) in units of the neighbourhood size $\epsilon$. For example, in the periodic case displayed in Fig.~\ref{periodic_list}, it takes $9$ iterations (time points) from vertex $1$ to $10$ in the time domain, while the shortest path to cover the phase space distance has only a length of $l_{1,10}=3$. Hence, shortest paths do not allow to infer the temporal evolution of the system. Even more, for the path concept in a recurrence network, no information about the temporal order of the individual observations is considered (for example, the shortest path between vertices $1$ and $2$ in Fig.~\ref{periodic_list} is given by the sequence (1,8,5,2), which is not ordered in time).

One should note that if the phase space is strongly fragmented (for instance, in the period-3 window of the logistic map, which has been discussed elsewhere \cite{Marwan2009}, the phase space consists of three discrete points), the resulting recurrence networks may be composed of different disconnected clusters. Furthermore, there might be more than one shortest path connecting two nodes. For example, in the aperiodic example in Fig.~\ref{nonperiodic_list}, the shortest path from node $1$ to node $7$, ($l_{1,7}=3$ as shown in Fig.~\ref{nonperiodic_list}(d)), can be obtained by three different choices, that are, (1,2,6,7), (1,5,3,7), and (1,5,6,7).

\subsubsection{Average path length.}

The average path length $\mathcal{L}$ is defined as the mean value of the shortest path lengths $l_{i,j}$ taken over all pairs of vertices $(i,j)$,
\begin{equation}
\mathcal{L} = \left<l_{i,j}\right>=\frac{2}{N(N-1)}\sum_{i<j} l_{i,j}.
\end{equation}
\noindent
Here, for disconnected pairs of vertices, the shortest path length is set to zero by definition. Note that in most practical applications, this has no major impact on the corresponding statistics.
 
The average phase space separation of states $\left<d_{i,j}\right>$ serves as an $\epsilon$-lower bound to $\mathcal{L}$, since 
\begin{equation}
d_{i,j} \leq \epsilon l_{i,j} \label{path_distance_inequality}
\end{equation}
\noindent 
due to the triangular inequality, and hence 
\begin{equation}
\left<d_{i,j}\right> \leq \epsilon \mathcal{L}.
\label{avgL_eq}
\end{equation}
\noindent 
Interpreted geometrically, this inequality holds because $\mathcal{L}$ approximates the average distance of states along geodesics on the recurrence network graph (which can be considered as the geometric backbone of the attractor) in multiples of $\epsilon$, while $\left<d_{i,j}\right>$ gives the mean distance of states in $\mathbb{R}^m$ as measured by the norm $\|\cdot\|$.

\subsubsection{Network diameter.}

By a similar argument as used in Eq. (\ref{path_distance_inequality}) for the average path length, the diameter 
\begin{equation}
\mathcal{D} = \max_{i,j} l_{i,j}
\end{equation}
\noindent 
of the recurrence network (i.e.~the maximum path length) serves as an $\epsilon$-upper bound to the estimated diameter 
\begin{equation}
\Delta = \max_{i,j} d_{i,j}
\end{equation}
\noindent 
of the attractor in phase space:
\begin{equation}
\Delta \leq \epsilon \mathcal{D}.
\end{equation}

\subsubsection{Closeness centrality.}\label{closeness}

The inverse average shortest path length of vertex $v$ to all others in the recurrence network is measured by the closeness centrality \cite{Freeman1977}
\begin{equation}
c_v = \frac{N-1}{\sum_{i=1}^N l_{v,i}}. \label{closeness_equation}
\end{equation}
\noindent 
If $i$ and $j$ are not connected, i.e., $A_{i,j}=0$, the maximum shortest path length in the graph, $N-1$, is used in the sum by definition. In a recurrence network, $c_v$ can be geometrically interpreted as measuring the closeness of $v$ to all other states with respect to the average length (in units of $\epsilon$) of geodesic connections on the recurrence network graph. In other words, $c_v$ is large if most of the other vertices are reachable in a small number of $\epsilon$-jumps from state to state.

From Eqs. (\ref{path_distance_inequality}) and (\ref{closeness_equation}), we can see that the inverse closeness $c_v^{-1}$ is bounded from below by the average phase space distance of vertex (state) $v$ to all other vertices (states) $\left<d_{v,i}\right>_i$ in units of $\epsilon$ (geometrical closeness), as measured by the norm $\|\cdot\|$,

\begin{equation}
\frac{1}{\epsilon} \frac{N}{N-1} \left<d_{v,i}\right>_i \leq c_v^{-1}.
\end{equation}

\noindent Put differently, geometrical closeness provides an upper bound for topological closeness,

\begin{equation}
c_v \leq \epsilon \frac{N-1}{N} \left<d_{v,i}\right>_i^{-1}.
\end{equation}

\subsubsection{Betweenness centrality.}\label{sec:bc}

The betweenness centrality $b_v$ has been originally introduced for characterising the importance of individual vertices for the transport of information or matter in general complex networks \cite{Freeman1977}. Unlike the degree centrality $k_v$, it is defined locally but depends on global adjacency information. 

Let us assume that information travels through the network on shortest paths. There are $\sigma_{i,j}$ shortest paths connecting two nodes $i$ and $j$. We then regard a node $v$ to be an important mediator for the information transport in the network, if it is traversed by a large number of all existing shortest paths. Betweenness centrality is given by
\begin{equation} 
b_v = \sum_{i,j\neq v}^N \frac{\sigma_{i,j}(v)}{\sigma_{i,j}}, 
\end{equation}
\noindent 
where $\sigma_{i,j}(v)$ gives the number of shortest paths from $i$ to $j$, that include $v$. Here the contribution of shortest paths is weighted by their respective multiplicity $\sigma_{i,j}$, the physical rational for this normalisation being that the total volume of information flow between two vertices, when summed over all shortest paths connecting them, should be the same for all pairs in the network. Hence, in addition to degree and closeness centralities, betweenness centrality yields another possibility to identify especially relevant vertices.

For a recurrence network, the notion of information transfer is not useful anymore. However, one may still argue in a geometric way that high betweenness states are typical for regions of sparse phase space density that separate different high-density clusters (refering to the information flow analogy mentioned above, one may consider the corresponding vertices as \textit{geometric bottlenecks}). Thus, the occurrence of high betweenness values can be a sign of highly fractionated attractors (on the scale resolved by the considered threshold $\epsilon$). A more detailed discussion of the corresponding implications for some simple model systems will be given in Section \ref{sec:examples}.

\subsubsection{Edge betweenness.}\label{sec:eb}

While betweenness centrality refers to vertex properties of a network, one may define an equivalent measure also based on the number of shortest paths on the network that include a specific edge $(i,j)$. We refer to the corresponding property as the edge betweenness $b_{i,j}$. Note that though there is a conceptual difference between vertex-related and edge-related betweenness, both quantities are indicators for regions of low phase space density that separate regions with higher density (or, to say it differently, of regions of high attractor fractionation) and thus have practically the same dynamical meaning.

\section{Examples}\label{sec:examples}

In the following, we will show the potentials of the network-theoretic measures discussed in the previous section for recurrence networks obtained from three paradigmatic chaotic model systems.

\subsection{Model systems}\label{sec:examples_models}

Basic results for one-dimensional maps have already been described for the logistic map (see \cite{Marwan2009}) based on numerical calculations and are supplemented by some further computations in the appendix. At this point, we prefer to discuss in some more detail the properties of systems that are defined in somewhat higher dimensions. In particular, we consider the H\'enon map
\begin{equation}
x_{i+1}=y_i+1-1.4x_i^2,\ y_{i+1}=0.3x_i
\end{equation}
\noindent
as an example for a chaotic two-dimensional map, and the R\"ossler system
\begin{equation}
\frac{d}{dt}\left(x,y,z\right)=\left(-y-z, x+0.2y, 0.2+z(x-5.7)\right)
\end{equation}
\noindent
as well as the Lorenz system
\begin{equation}
\frac{d}{dt}\left(x,y,z\right)=\left(10(y-x), x(28-z)-y, xy-\frac{8}{3}z\right)
\end{equation}
\noindent
as two examples for three-dimensional chaotic oscillators. In all following considerations, no additional embedding will be used. Note, however, that for the continuous systems, temporal correlations between subsequent observations have been excluded by removing all sojourn points \cite{Gao1999}.

\begin{figure}
  \centering
  \includegraphics[scale=0.5]{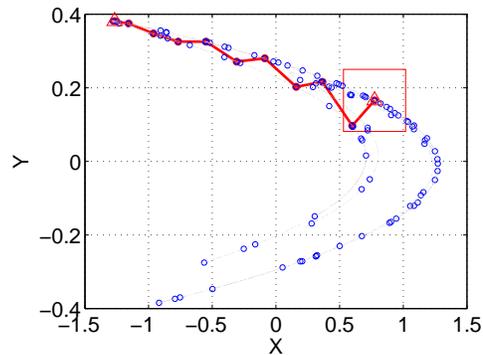}
\caption{\small One trajectory of the H\'enon map (attractor indicated by grey dots) formed by 100 iterations (i.e.~$N=101$), indicated by circles. The initial condition is marked by a square the size of which corresponds to the neighbourhood threshold $\epsilon=0.25$ (maximum norm) considered in the derivation of the corresponding recurrence network. The shortest path (with $l_{1,101}=10$) between the first and the last point is indicated by a continuous red line.}
\label{p2p_henon}
\end{figure}

\begin{figure*}
  \centering
  \includegraphics[scale=0.5]{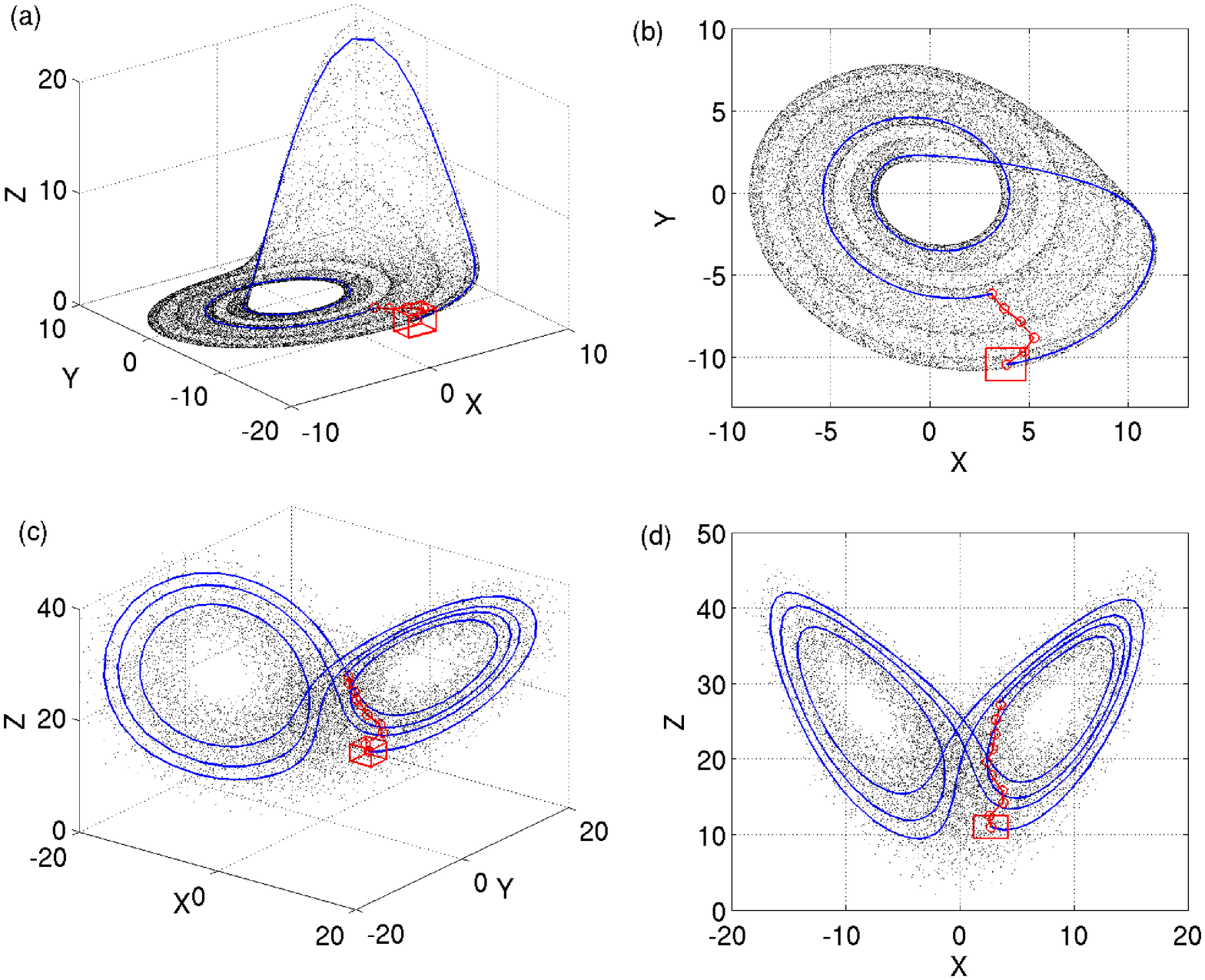}
\caption{\small (a,b) Chaotic attractor of the R\"ossler system (dark dots) and realisation of one particular non-periodic trajectory (blue line), corresponding to $T=N\Delta t=11.4$ time steps. The size of the considered neighbourhood $\epsilon=1$ (maximum norm) is indicated by a red square around the initial condition. (Note that the square gives an idealised representation of the considered neighbourhood.) The red line displays the shortest path between the initial condition and the final value on this trajectory ($l=5$). (c,d) One example trajectory of the Lorenz system with $T=5$ time steps, and resulting shortest path between initial and final state ($\epsilon=1.5$, $l=9$).}
\label{p2p_rossler}
\end{figure*}

Figs.~\ref{p2p_henon} and \ref{p2p_rossler} show examples of typical trajectories of these three model systems. In addition, the shortest paths between the first and last point of the individual realisation are indicated, underlining the deep conceptual differences between the concepts of trajectory (in phase space) and path (in a recurrence network, see Section \ref{sec:recnet}).


\subsection{$\epsilon$-dependence of global network measures}

Let us first consider the dependence of the global network measures $\mathcal{L}$, $\mathcal{C}$ and $\mathcal{R}$ on the choice of the threshold $\epsilon$ for our three model systems.

\begin{figure*}
  \centering
  \includegraphics[scale=0.5]{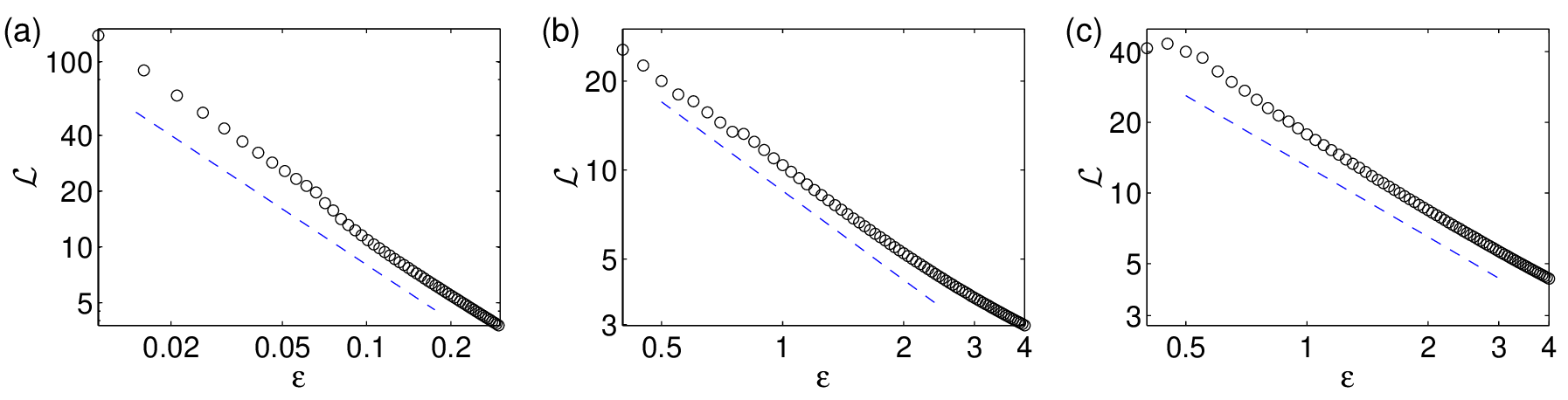} \\
  \includegraphics[scale=0.5]{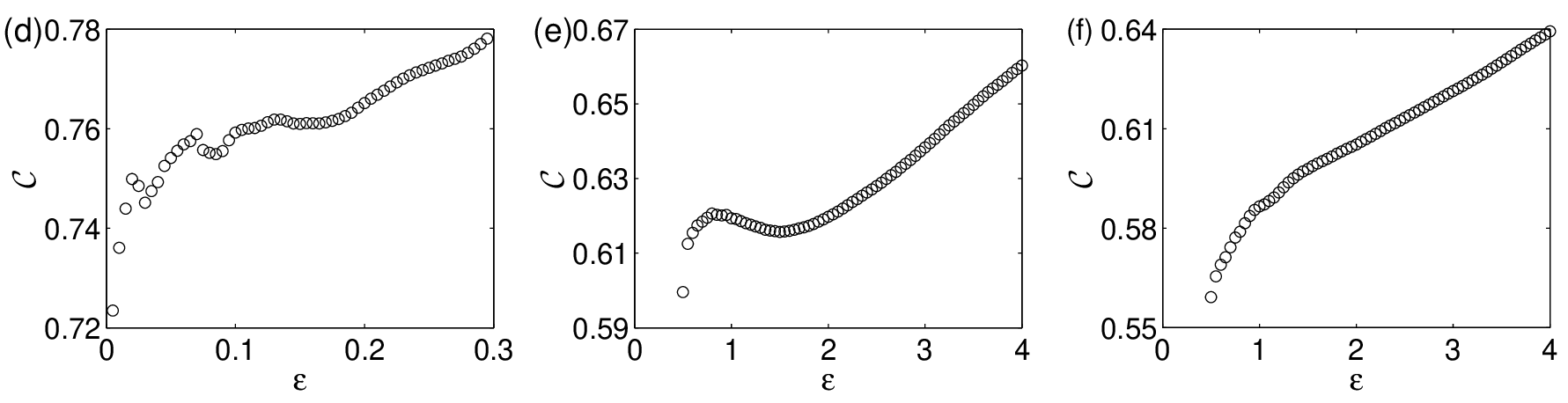}  \\
  \includegraphics[scale=0.5]{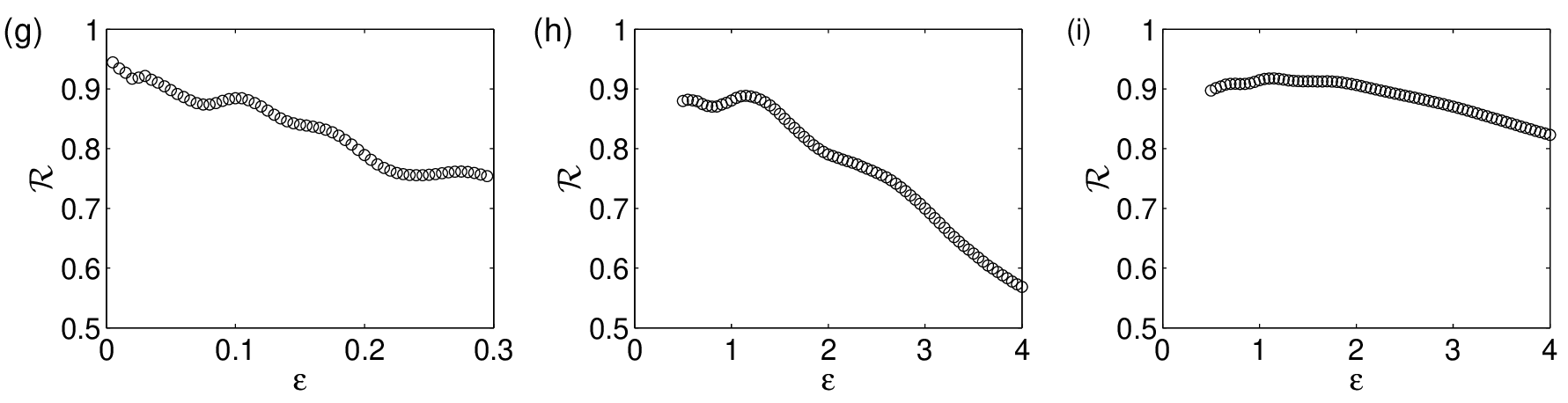} 
\caption{\small Dependence of (a), (b), (c) the average path length $\mathcal{L}$, (d), (e), (f) the global clustering coefficient $\mathcal{C}$, and (g), (h), (i) the assortativity coefficient $\mathcal{R}$ for the H\'enon map, R\"ossler, and Lorenz system (from left to right) using the maximum norm. Dashed lines in the plots on $\mathcal{L}(\epsilon)$ indicate the approximate presence of the theoretically expected $1/\epsilon$ dependence of the average path length. Note that although a normalised threshold $\epsilon$ might yield a better comparability of the results for the different systems, we prefer using the absolute values here since the typical normalisation factors -- either an empirical estimate of the standard deviation of the phase space density (which may be crucially influenced by asymmetric densities) or the attractor diameter (which is itself not a priori known in advance) have certain conceptual problems.
} 
\label{avg_path_all}
\end{figure*}

The variations of $\mathcal{L}$ with the threshold $\epsilon$ are shown in Fig.~\ref{avg_path_all} and verify the existence of an inverse relationship of a corresponding lower bound postulated in Eq.~(\ref{avgL_eq}).

For the global clustering coefficient, the dependence on $\epsilon$ is more complicated and depends on the specific properties of the considered system (Fig.~\ref{avg_path_all}). In particular, while for too small $\epsilon$, problems may occur, since the recurrence network may decompose into different disconnected clusters for a length $N$ of the considered time series, for intermediate threshold values, an approximately linear increase of $\mathcal{C}$ with $\epsilon$ seems to be a common feature of all three examples. Following the discussion of the behaviour of one-dimensional maps in \ref{app_1dmap}, we may argue that this increase is most likely related to the effect of the attractor boundaries.

Finally, concerning the assortativity coefficient $\mathcal{R}$, we observe that for small $\epsilon$, the recurrence networks are highly assortative (e.g.~$\mathcal{R}$ is close to 1). This behaviour can be related to the fact that in case of small neighbourhoods, these phase space regions are usually characterised by only weak variations of the phase space density, so that neighbouring vertices have a tendency to obey a similar degree. As $\epsilon$ becomes larger, larger regions of the phase space are covered, where the density may vary much stronger, which implies that the degrees of neighbouring vertices become less similar. Note, however, that since in this case, the mutual overlap of the different neighbourhoods becomes successively larger, there is still a significantly positive correlation between the degrees of neighbouring vertices. One may further observe that the decrease of $\mathcal{R}$ with $\epsilon$ may be interrupted by intermediate increases, which are probably related to some preferred spatial scale of the separation of certain dynamically invariant objects such as unstable periodic orbits (UPOs). We will come back to this point in Sec.~\ref{cluster_spatial}.

\subsection{Spatial distributions of vertex properties}\label{sec:examples_spatial}

In the following, we will study the interrelationships between local network properties and structural features of the phase space for the three considered chaotic model systems.

\subsubsection{Degree centrality.} 

\begin{figure*}[htbp]
  \centering
  \includegraphics[width=\textwidth]{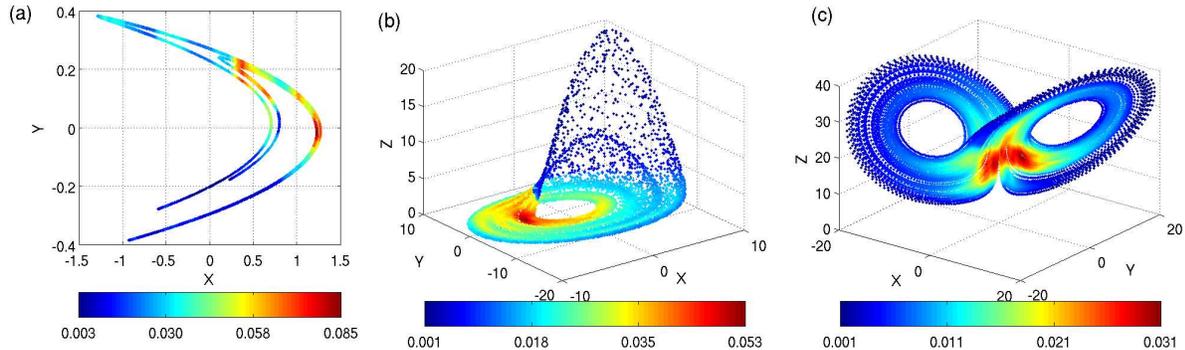}
\caption{\small Colour-coded representation of the local recurrence rate $RR_v$ (proportional to the degree centrality $k_v$) in phase space (a) H\'enon map ($N=10,000$), (b) R\"ossler ($N=10,000$), and (c) Lorenz systems ($N=20,000$). The value of $\epsilon$ for each case is chosen in such a way that the global recurrence rate $RR = \rho \approx 0.03$.  
\label{degree_all} }
\end{figure*}

When considering the degree centrality $k_v$ or, equivalently, the local density $\rho_v$ for all vertices of the network, a broad range of variability is found (Fig.~\ref{degree_all}). In particular, the behaviour follows the expectation that regions with a high phase space density (for example, the merger of the two scrolls of the Lorenz oscillator) also reveal a high density of vertices and, hence, high degree centralities. Note that the calculation of a recurrence plot depends on the parameter $\epsilon$, which should be tailored to the considered system under study and the specific questions one wishes to address. Several "rules of thumb" for the choice of the threshold $\epsilon$ have been advocated in the literature~\cite{Marwan2007,Schinkel2008}. It has been suggested that the choice of $\epsilon$ to achieve a fixed recurrence rate $RR$ is helpful for the estimation of dynamical invariants in many systems \cite{Marwan2007}. Therefore, this procedure will be adopted here to obtain an overall visualisation of the degree centrality $k_v$ in phase space, with $RR = \rho \approx 0.03$ (which lies within the typical scaling region of the correlation integral). However, as we will see later, for the local clustering coefficients (Sec.~\ref{cluster_spatial}) disclosing local fine structures of the phase space density, it is necessary to choose smaller $\epsilon$.  

\subsubsection{Closeness centrality.}

\begin{figure}
  \centering
    \includegraphics[width=\textwidth]{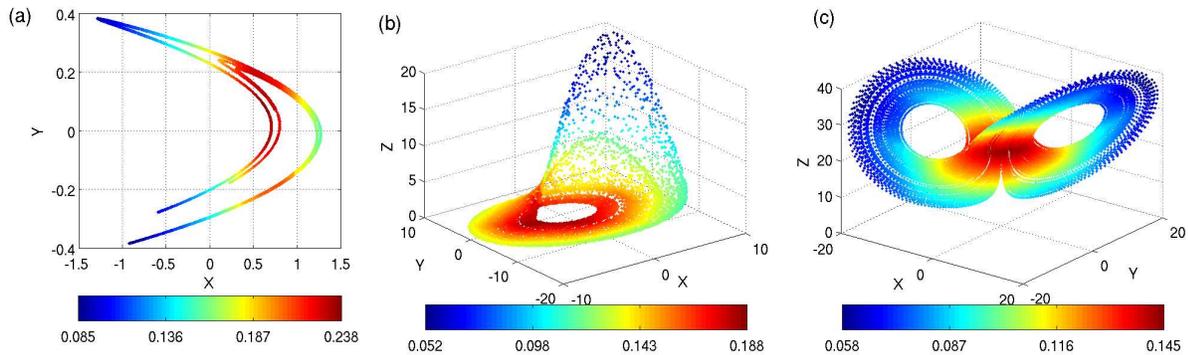}
\caption{\small Colour-coded representation of the closeness centrality $c_v$ (Eq.~(\ref{closeness_equation})) in
phase space (a) H\'enon map, (b) R\"ossler, and (c) Lorenz systems ($N$ and $\epsilon$ as in Fig.~\ref{degree_all}).}
\label{closeness_all}
\end{figure}

Figure~\ref{closeness_all} reflects the spatial distribution of the closeness centrality $c_v$ . In good agreement with our previous theoretical considerations on the geometric meaning of this measure (Sec.~\ref{closeness}), we find high values of $c_v$ near the centre of gravity of the attractor in phase space, and low values at phase space regions that have large distances from this centre.

\subsubsection{Clustering coefficient.} \label{cluster_spatial}

\begin{figure}
  \centering
  \includegraphics[scale=0.5]{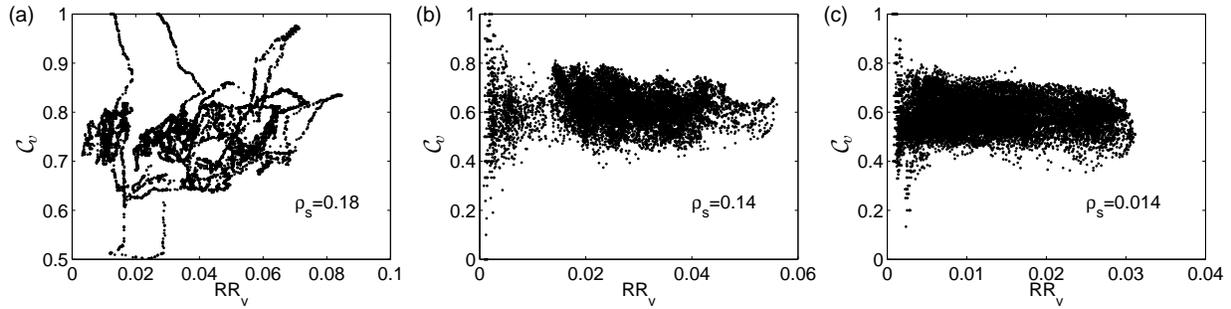}
\caption{\small Scatter diagrams between the local recurrence rate $RR_v$ and the local clustering coefficient $\mathcal{C}_v$ for typical realisations of (a) H\'enon map, (b) R\"ossler, and (c) Lorenz system. The inserted values $\rho_s$ give the corresponding rank-order correlation coefficient (Spearman's Rho).} \label{kvcv_all}
\end{figure}

\begin{figure*}
  \centering
    \includegraphics[scale=0.5]{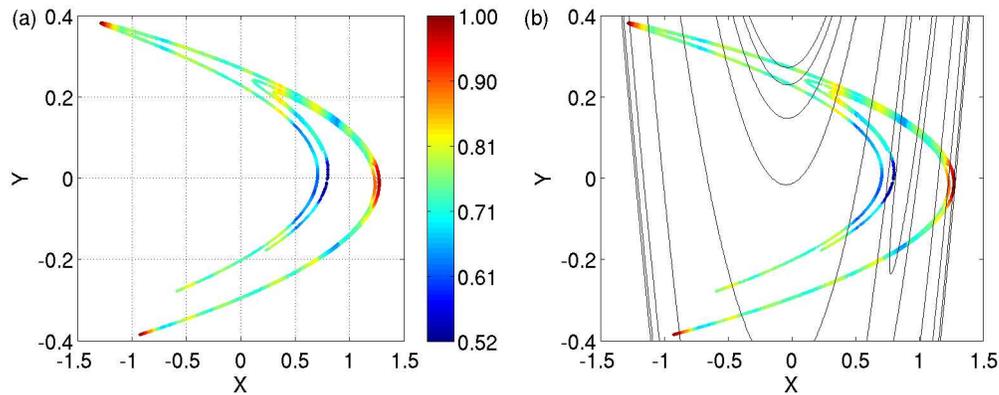}
\caption{\small (a) Colour-coded representation of the local clustering coefficients $\mathcal{C}_v$ for the H\'enon map in its phase space. (b) Relationship between the local clustering coefficients and the stable manifold of the H\'enon map. The finite length segment of the stable manifold is calculated by the method described in~\cite{Ott1993,Nusse1994} (with 20,000 iterations).}
\label{cc_henon}
\end{figure*}

\begin{figure*}
  \centering
    \includegraphics[scale=0.5]{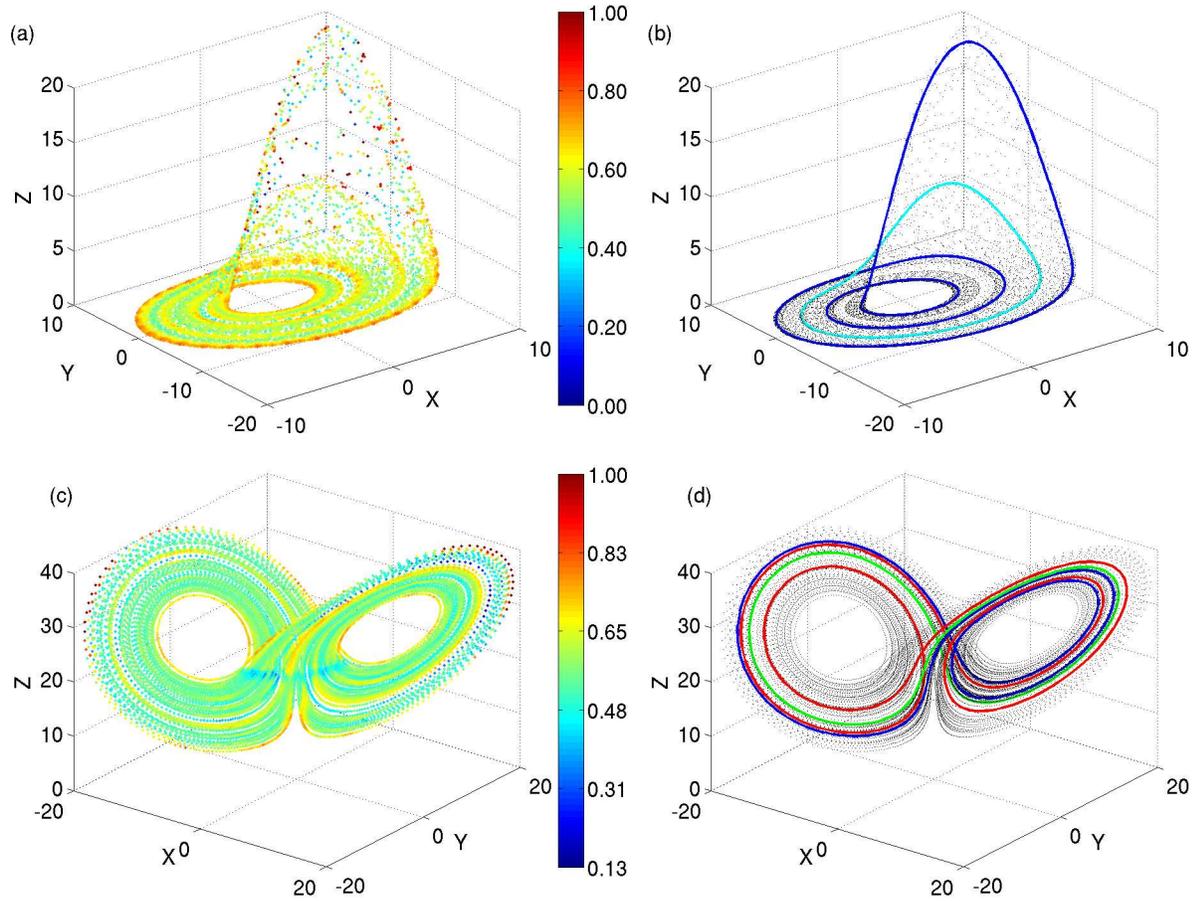}
\caption{\small (a) Colour-coded representation of the local clustering coefficients $\mathcal{C}_v$ for the R\"ossler system ($N=10,000$) in its phase space. (b) Several locations of the unstable periodic orbits with low periods, obtained using a method based on the windows of parallel lines in the corresponding recurrence plot as described in~\cite{Marwan2007}. The chosen value of $\epsilon$ corresponds to a global recurrence rate of $RR=0.01$. (c),(d) Same as (a),(b) for the Lorenz system ($N=20,000$).}
\label{cc_ros}
\end{figure*}

Concerning the local clustering coefficient, one may suppose that in the case of high-density regions in phase space, there are many vertices located in the vicinity of a specific vertex, in particular, in a distance that does not exceed $\epsilon/2$. By definition, these vertices must then also be adjacent to each other, which gives considerable contributions to the clustering coefficient. In contrast, for low-density regions, one may argue that even if there are more than one vertices in some $\epsilon$-neighbourhood of a vertex (i.e., $k_v\geq 2$), it is less likely that these are also separated by a distance that is smaller than $\epsilon$. Following these considerations, one might expect some relationship between the degree centrality and the local clustering coefficient. However, as Figs.~\ref{kvcv_all},~\ref{cc_henon}, and \ref{cc_ros} demonstrate, $\mathcal{C}_v$ does 
clearly reveal more and different structural properties than the degree centrality alone (note that here, a smaller value of $\epsilon$ has been chosen to disclose the local fine structures of the phase space density). In particular, a visual comparison with Fig.~\ref{degree_all} reveals that the clustering coefficient characterises some specific higher-order characteristics of the phase space density.  

Beside the effect of the local phase space density, we argue that the local clustering coefficient does also depend on the spatial filling (i.e., the homogeneity of the phase space density) in the neighbourhood of the considered point. In particular, in the case of a two-dimensional system, an alignment of vertices along a one-dimensional subspace will produce a clearly lower clustering coefficient than a homogeneous filling of the neighbourhood. This behaviour is underlined in Fig.~\ref{cc_henon} for the H\'enon map, where maximum values of $\mathcal{C}_v=1$ can be particularly found at the two tips of the attractor. Hence, the local clustering coefficient can be considered as an entropy-like characteristic in that it quantifies the homogeneity of the phase space density in the neighbourhood of a vertex. From the theory of spatial random graphs \cite{Dall2002}, which may be assumed to yield the lowest possible clustering coefficients among all spatial networks, it is known that for a given dimension $m$ of the considered system, in the asymptotic limit $N\to\infty$ and $\epsilon\to 0$, the possible values of $\mathcal{C}_v$ are bound between the corresponding theoretical value and $1$. Note that this lower bound systematically decreases with increasing $m$, which appears to be reasonable if one interpretes $\mathcal{C}_v$ as an entropy-like quantity. To be more specific, according to Dall and Christensen \cite{Dall2002}, this decay follows an exponential function for sufficiently large embedding dimensions.

The presence of distinct structures in the spatial profile of the local clustering coefficient is related to the emergence of specific dynamically invariant objects in the considered model systems. In the case of H\'enon map (Fig.~\ref{cc_henon}), there is a clear tendency that points that are close to the stable manifold associated with the system have remarkably higher values of $\mathcal{C}_v$. Note, however, that because of finite size effects, this coincidence cannot be found for all corresponding regions of the phase space. For the two continuous systems (Fig.~\ref{cc_ros}), points close to the trapping regimes of UPOs have higher clustering coefficients. It is, in some sense, trivial to understand the role of UPOs in forming such regimes of higher clustering. Whenever the trajectory of the corresponding systems visits the neighbourhood of an UPO, it is captured in this neighbourhood for a certain finite time, during which the probability of recurrences is increased. Furthermore, once the trajectory is trapped, the local divergence rate becomes smaller. This smaller local divergence rate is captured by the clustering coefficient (in terms of higher-order correlations between neighbours of a vertex). As for the finite-$\epsilon$ effect in the H\'enon map, the regions with increased clustering coefficients in most cases only coincide with UPOs of lower periods. Therefore, in Fig.~\ref{cc_ros}, only a few UPOs of low order are shown for comparison. Note that if two UPOs are separated by a distance smaller than $\epsilon$ in phase space, the clustering coefficient is not able to distinguish between these two structures and, hence, shows a broad band with increased values. Following this argumentation, $\mathcal{C}_v$ is a useful measure for detecting phase space regions with a high density of low-order UPOs, which is in good agreement with corresponding considerations in \cite{Gao2009}.

\subsubsection{Betweenness centrality.}

\begin{figure*}
  \centering
    \includegraphics[scale=0.5]{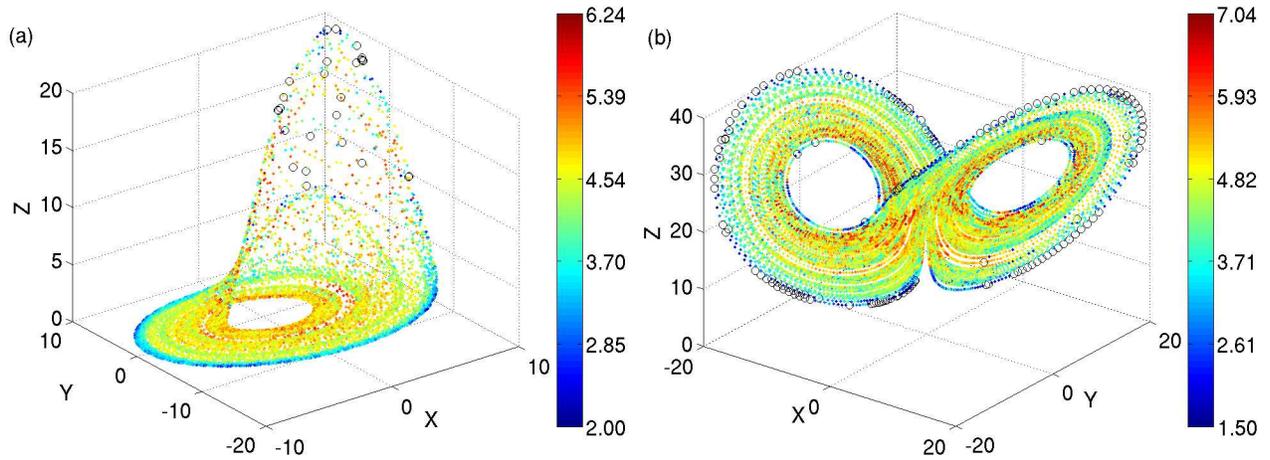}
\caption{\small Logarithm of the betweenness centrality $b_v$ for (a) R\"ossler and (b) Lorenz system ($N$ and $\epsilon$ as in Fig.~\ref{cc_ros}). Points shown as circles have betweenness values below the lower limit of the displayed colour scale.} 
\label{bc_ros}
\end{figure*}

Our interpretation of the betweenness centrality in Sec.~\ref{sec:bc} implies that $b_v$ is a rather sensitive measure of the local fragmentation of the attractor and thus may give complementary information especially on very small scales. Unfortunately, numerical limitations in the calculation of this measure did not allow us to explore the limit of small neighbourhoods ($\epsilon\to 0$). However, from our computations with somewhat larger thresholds (see Fig.~\ref{bc_ros}), we can already derive some general statements about the behaviour of betweenness centrality for the considered model systems. First, note that regions close to the outer boundaries of the attractor (in contrast to those in the vicinity of the inner boundaries, e.g., of the R\"ossler oscillator) are not important for many shortest path connections on the recurrence network. Hence, vertices settled in the corresponding parts of the phase space are characterised by low betweenness values. Second, if there are pronounced regions with rather few isolated points in between high-density regions (for example, between two UPOs in the R\"ossler or Lorenz systems), there is an increasing number of shortest paths crossing these vertices, which leads to higher values of $b_v$. In turn, vertices in the vicinity of UPOs (i.e., high-density regions) show lower betweenness values. Therefore, betweenness centrality provides a complementary view on the attractor geometry in comparison to the local clustering coefficient $\mathcal{C}_v$ (Fig.~\ref{cc_ros}). 

\subsection{Spatial distributions of edge properties}\label{sec:examples_spatial2}

\begin{figure*}
  \centering
  \includegraphics[scale=0.5]{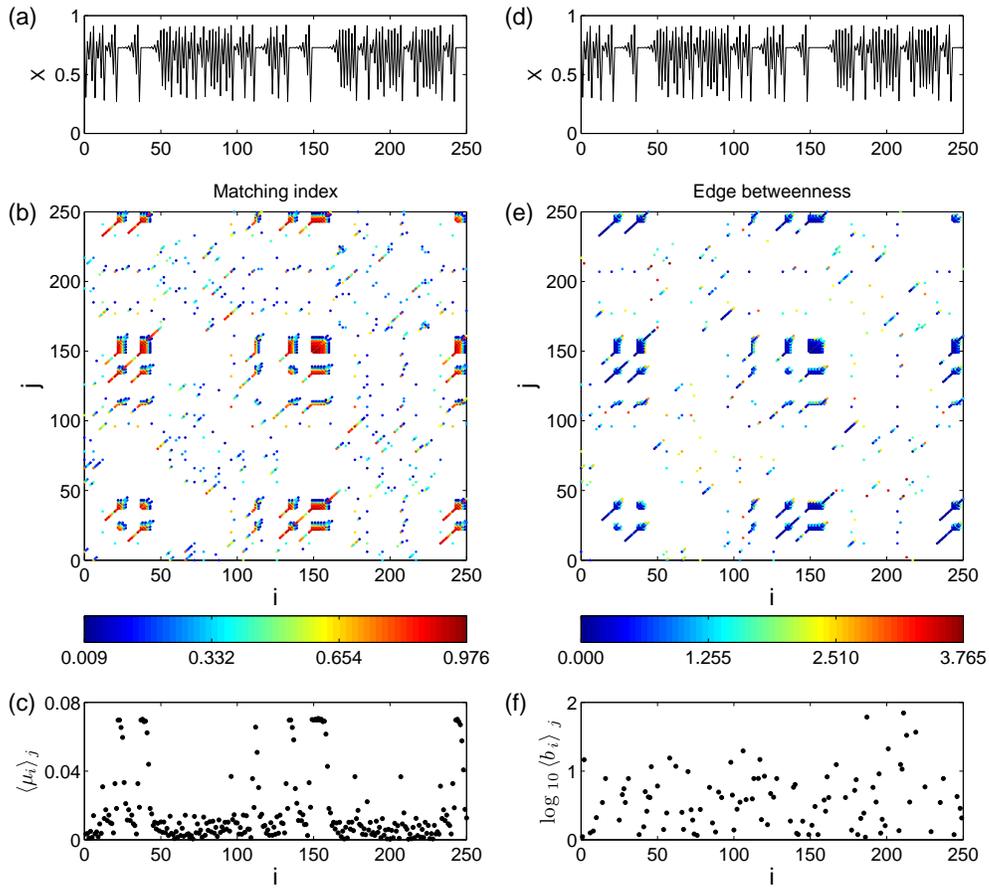}
\caption{\small (a,d) One example trajectory of the logistic map at $a=3.679$ (chaotic regime with intermittency), $\epsilon=0.015\sigma_x$, $N=1000$ (only a part of the trajectory with $250$ points is shown). (b) Matching index $\mu_{i,j}$ between all pairs of vertices (colour-coded). (c) Average matching index $\mu_i$ of all vertices of the considered recurrence network. (e,f) As in (b,c) for the logarithm of the edge betweenness $b_{i,j}$.}
\label{mi_eb}
\end{figure*}

Similarly to the local vertex properties, one may also study the characteristics of different edges in the recurrence networks. Since the resulting structures are more pronounced than for the three model systems considered so far, Fig.~\ref{mi_eb} shows the matching index and edge betweenness for one realisation of the logistic map $x_{i+1}=ax_i(1-x_i)$ in the intermittent chaotic regime (see \cite{Marwan2009}). The presence of intermittent dynamics can be clearly seen from the recurrence plots in terms of extended square recurrence patterns, which hence lead to mutually connected vertices of the associated recurrence networks that correspond to subsequent points in time.

\begin{figure*}
  \centering
  \includegraphics[scale=0.5]{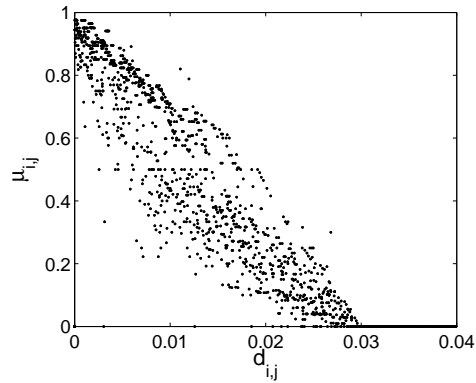}
\caption{\small Scatter diagram of the matching index $\mu_{i,j}$ against the phase space distance $d_{i,j}$ for the logistic map at $a=3.679$ (parameters as in Fig.~\ref{mi_eb}).}
\label{mi_dist}
\end{figure*}

Figures~\ref{mi_eb} and \ref{mi_dist} show the complex dependence between phase space distance $d_{i,j}$, matching index $\mu_{i,j}$, and edge betweenness $b_{i,j}$. For the matching index, the results are consistent with our theoretical considerations presented in Sec.~\ref{sec:mi}. In particular, for $d_{i,j}\to 0$, we have $\mu_{i,j}\to 1$, while for $d_{i,j}\to 2\epsilon$, $\mu_{i,j}\to 0$. Concerning the temporal evolution during the laminar (intermittent) phase, one may recognise that at the beginning, there is hardly any change in the state of the system, hence, $d_{i,j}$ is very small for subsequent points in time (vertices of the recurrence network), which relates to large values of the matching index near $1$. As the laminar phases are close to their termination, chaotic variations emerge and rise in amplitude, which leads to a subsequent increase of $d_{i,j}$ and, hence, decrease of $\mu_{i,j}$. 

Concerning the edge betweenness $b_{i,j}$ (the spatial pattern of which is very similar to that of the vertex-based betweenness centrality $b_v$ due to the spatial proximity of edge and corresponding vertices), the overall behaviour is opposite to that of $\mu_{i,j}$. During laminar phases, we find that since all states are very close to each other, possible shortest connections may alternatively pass through a variety of different edges, leading to low values of the edge betweenness. Close to the termination, there is in turn an increase of this measure. However, the most interesting feature of the edge betweenness is presented by isolated edges with very high values of $b_{i,j}$, which correspond to rarely visited phase space regions between intervals of higher phase space density. More specifically, the average edge betweenness of vertices in such low-density regions may exceed that of high-density regions by orders of magnitude (Fig.~\ref{mi_eb}).


\section{Conclusions}

This paper has reconsidered the analysis of time series from complex systems by means of complex network theory. We have argued that most existing approaches for such an analysis suffer from certain methodological limitations or a lack of generality in their applicability. As an appealing solution, we have suggested recurrence networks as a unifying framework for studying time series as complex networks, which is based on the idea of recurrence plots. As we have argued, this specific approach is applicable to univariate as well as multivariate time series without or with embedding. In addition, recurrence networks can be applied for studying time series with non-equidistant time-scales and allow the construction of simple significance tests with respect to the associated network-theoretic measures \cite{Marwan2009}.

As a main achievement, we have provided a thorough reinterpretation of a variety of statistical measures from network theory computed for recurrence networks in terms of phase space properties of dynamical systems. Since all time ordering information is lost in this approach, all complex network characteristics are dynamically invariant, i.e., they are only sensitive to certain properties of the invariant density of the considered dynamical system. From this invariance, it follows that specific measures such as the local clustering coefficient may be used for detecting dynamically invariant objects like unstable periodic orbits or chaotic saddles. On the other hand, this feature also implies that the proposed method cannot be used to distinguish between deterministic (chaotic) and stochastic systems, which is exemplified by our comparison between Bernoulli map and uniform noise in \ref{app_1dmap}. As a possibility to overcome this potential point of criticism to our method, we emphasise that an additional embedding should change the properties of deterministic systems in a different way than for a stochastic system, hence, studying complex network properties in dependence on the embedding dimension might help solving this interpretation problem. We will further elaborate this idea in future research.

Using wide-spread statistical characteristics of complex networks such as the ``trinity'' of centrality measures (degree, closeness, betweenness) and the clustering coefficient, we were able to provide a detailed interpretation of the corresponding results for recurrence networks in terms of higher-order phase space properties. In particular, degree centrality relates to the local density, closeness centrality to the average geometrical proximity of an observation to all other observations, and betweenness centrality to the local fragmentation of points in phase space. For the clustering coefficient, the consideration of different model systems has revealed that apart from possible density effects, the clustering properties are related to the spatial filling of the phase space, which become important close to the attractor boundaries and certain dynamically invariant objects such as invariant manifolds or unstable periodic orbits. Additionally, we have presented a rigorous analytical treatment of the local and global clustering coefficients of the recurrence networks of one-dimensional chaotic maps, which perfectly matches our numerical results (see \ref{app_1dmap}). A possible further relationship between the local clustering coefficient of a recurrence network and the local Lyapunov exponent of the underlying dynamical system remains a topic for future work.

With respect to existing recurrence plot based methods of time series analysis, e.g., recurrence quantification analysis (RQA), we would like to emphasise that our approach yields a complementary view on the phase space properties of the underlying dynamical system. In particular, one should note that nearly all of the considered network-theoretical measures have no direct equivalents in traditional RQA and vice versa. Hence, one may think of situations when either one of the two frameworks (i.e., RQA or recurrence networks) may provide better results than the other. In turn, there have been some recent approaches on applying methods of time series analysis to general complex networks (e.g.~\cite{Yang2004,Li2008}). We would like to underline that due to the duality between the recurrence matrix of a time series and the adjacency matrix of the associated recurrence network, RQA might be another promising candidate for this purpose (as long as one restricts oneself to measures that are invariant under re-ordering of time), which might yield interesting complementary insights on complex networks in a variety of different situations. A more detailed investigation of a corresponding approach will be subject of future research.

\vspace{0.3cm}
\textbf{Acknowledgements.} This work was partly supported by the German Research Foundation (DFG project no.~He~2789/8-2 and SFB 555 project C1) and the Japanese Ministry for Science and Education.

\appendix

\section{Clustering coefficient of recurrence networks for one-dimensional maps} \label{app_1dmap}

In order to compute the clustering coefficient of the recurrence network of a dynamical system, certain system-specific integrals have to be solved. For the case of one-dimensional maps that are defined on the integral $[0,1]$, these integrals can be explicitly expressed and eventually also evaluated analytically.

\subsection{General treatment}

\begin{figure*}
\centering
\resizebox{0.7\textwidth}{!}{\includegraphics{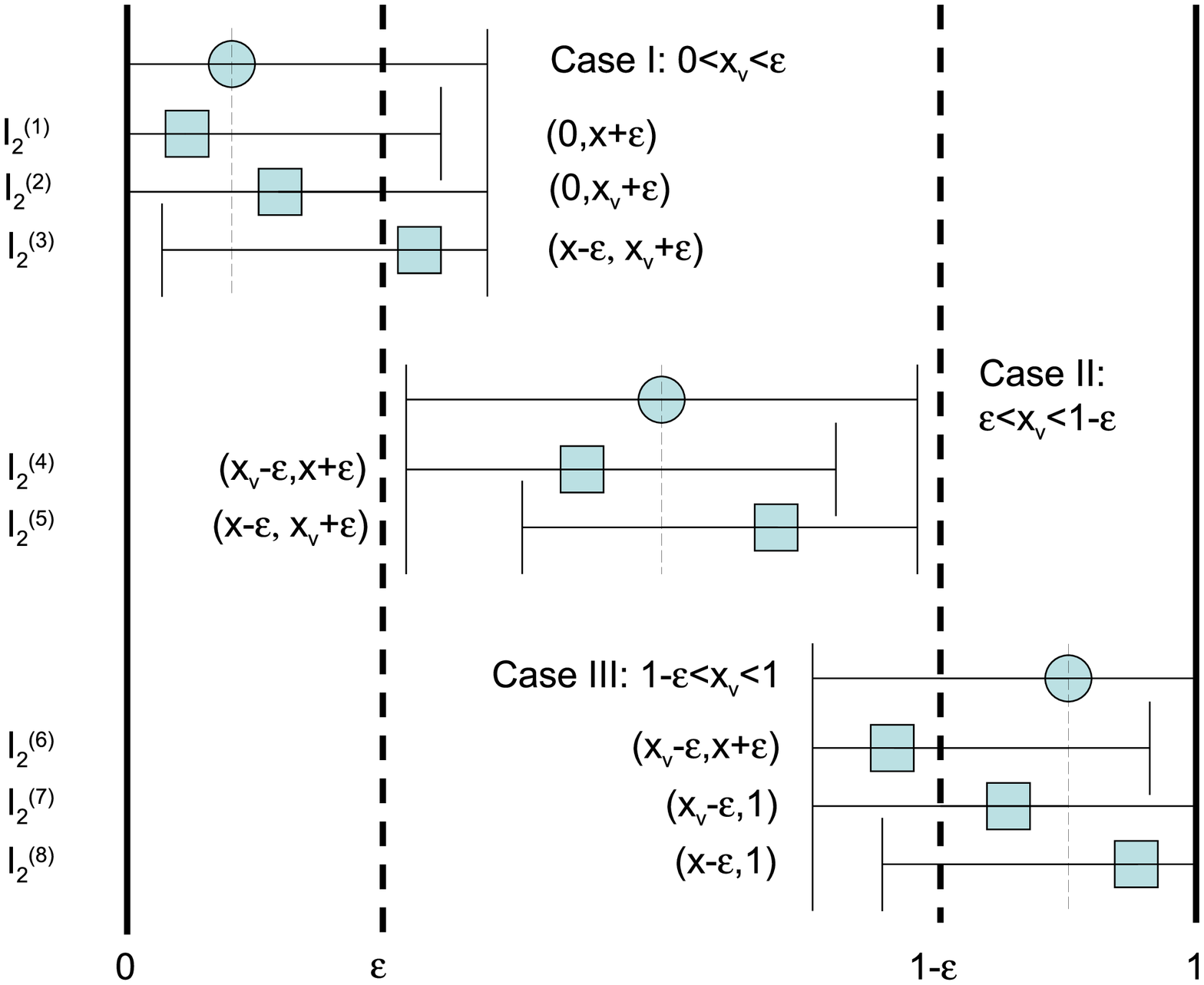}}
\caption{\small Schematic representation of the integration boundaries entering the three-point relationships in Eq.~(\ref{threepoint}) for $\epsilon\leq 0.5$.} 
\label{clust_1d}
\end{figure*}

The computation of the local clustering coefficient involves the computation of certain integrals to estimate the required probability terms discussed in Section \ref{sec:recnet}. In particular, we have that for a given vertex $v$ at a point $x_v$,
\begin{equation}
\begin{split}
P(A_{v,i}=1,A_{v,j}=1) &= P(||x_v-x_i||<\epsilon,||x_v-x_j||<\epsilon) \\
&= P(||x_v-x_i||<\epsilon)P(||x_v-x_j||<\epsilon) \\
&= P(||x_v-x_i||<\epsilon)^2 = \left[ \int_{x_v-\epsilon}^{x_v+\epsilon} dx\ p(x) \right]^2
\end{split}
\end{equation}
\noindent
for the condition in the denumerator of Eq.~(\ref{cc_ws}), where we have made use of the fact that the observations $x_i$ and $x_j$ can be considered to be independent of each other. However, this independence condition is not fulfilled in the corresponding numerator $P(A_{ij}=1,A_{vi}=1,A_{vj}=1)$, which therefore requires a subtle choice of the integration boundaries to assure a correct treatment of the three-point relationships
\begin{equation}
P(A_{i,j}=1,A_{v,i}=1,A_{v,j}=1) = \int_{x_v-\epsilon}^{x_v+\epsilon} dx\ p(x) \int_{\max(x-\epsilon,x_v-\epsilon)}^{\min(x+\epsilon,x_v+\epsilon)} dy\ p(y).
\label{threepoint}
\end{equation}
\noindent
As one may see in Fig.~\ref{clust_1d}, the correct integration is not trivial. In particular, if we use the abbreviations
\begin{eqnarray}
I_1(a,b) &=& \int_a^b dx\ p(x) \\
I_2(a,b;c,d) &=& \int_a^b dx\ \left[ p(x) \int_c^d dy\ p(y) \right]
\end{eqnarray}
\noindent
we obtain the following expressions for the local clustering coefficient $\mathcal{C}_v=\mathcal{C}(x_v)$ if $\epsilon\leq 0.5$:\\

\noindent
\underline{$0\leq x_v \leq\epsilon$:}
\begin{equation}
\begin{split}
\mathcal{C}_v&=\frac{I_2(0,x_v;0,x+\epsilon)+I_2(x_v,\epsilon;0,x_v+\epsilon)+I_2(\epsilon,x_v+\epsilon;x-\epsilon,x_v+\epsilon)}{I_1(0,x_v+\epsilon)^2}\\
&=\frac{I_2^{(1)}+I_2^{(2)}+I_2^{(3)}}{I_1(0,x_v+\epsilon)^2},
\end{split}
\end{equation}

\noindent
\underline{$\epsilon\leq x_v\leq 1-\epsilon$:}
\begin{equation}
\begin{split}
\mathcal{C}_v&=\frac{I_2(x_v-\epsilon,x_v;x_v-\epsilon,x+\epsilon)+I_2(x_v,x_v+\epsilon;x-\epsilon,x_v+\epsilon)}{I_1(x_v-\epsilon,x_v+\epsilon)^2}\\
&=\frac{I_2^{(4)}+I_2^{(5)}}{I_1(x_v-\epsilon,x_v+\epsilon)^2},
\end{split}
\end{equation}

\noindent
\underline{$1-\epsilon\leq x_v\leq 1$:}
\begin{equation}
\begin{split}
\mathcal{C}_v&=\frac{I_2(x_v-\epsilon,1-\epsilon;x_v-\epsilon,x+\epsilon)+I_2(1-\epsilon,x_v;x_v-\epsilon,1)+I_2(x_v,1;x-\epsilon,1)}{I_1(x_v-\epsilon,1)^2}\\
&=\frac{I_2^{(6)}+I_2^{(7)}+I_2^{(8)}}{I_1(x_v-\epsilon,1)^2}.
\end{split}
\end{equation}

Understanding the global clustering coefficient as the expectation value of the local one (taken over the whole possible range in $x$), one may use the following expression for deriving its value:
\begin{equation} \label{cc_global}
\mathcal{C}(\epsilon)=\int_{0}^{1} dx_v\ p(x_v)\ \mathcal{C}_v(x_v;\epsilon) = \int_{0}^{1} dx_v\ p(x_v)\ \frac{P(A_{ij}=1,A_{vi}=1,A_{vj}=1)}{P(A_{vi}=1,A_{vj}=1)}.
\end{equation}

\subsection{Bernoulli map}

\begin{figure*}
  \centering
  \includegraphics[scale=0.5]{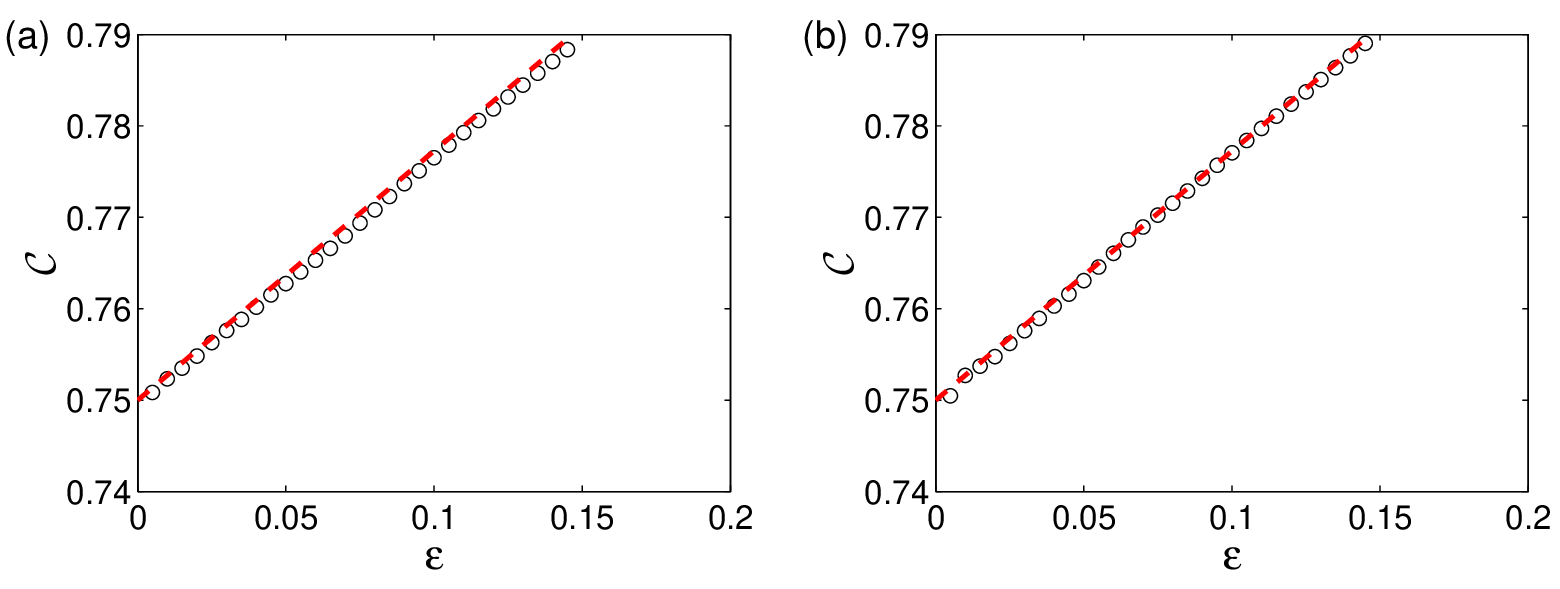}
\caption{\small Dependence of the local clustering coefficient $\mathcal{C}_v$ on the threshold $\epsilon$ for (a) the Bernoulli map and (b) uniformly distributed noise with values in $[0,1]$. Note that due to the same distribution of the data, i.e., the same density in the one-dimensional phase space, both curves are equal and match the theoretical expectations (dashed red lines).} \label{ccav_bernoulli}
\end{figure*}

Among all nonlinear maps defined on the unit interval $[0,1]$, the Bernoulli map $x_{i+1}=2x_i \mod 1$ has the simplest possible invariant density $p(x)\equiv 1$\footnote{Note that from the recurrence network properties, this map cannot be distinguished from a stochastic process with a uniform density in the same interval (see Fig.~\ref{ccav_bernoulli}).}. This allows an easy evaluation of the integrals for an analytic computation of the local clustering coefficient, being aware of the restricted integration range. As a result, one finds:
\begin{equation}
\mathcal{C}_v(x_v;\epsilon)=\left\{
\begin{array}{ll}
1-\left(\frac{x}{x+\epsilon}\right)^2, & 0\leq\ x\leq \epsilon \\
\frac{3}{4}, & \epsilon\leq x\leq 1-\epsilon \\
1-\left(\frac{1-x}{1-x+\epsilon}\right)^2, & 1-\epsilon\leq x\leq 1 \end{array} \right. .
\end{equation}
\noindent
In particular, for $\epsilon\to 0$ (and $N\to\infty$), we have $C_v\to \frac{3}{4}$ $\forall\ x\in ]0,1[$, which corresponds to the value of random geometric graphs in one dimension \cite{Dall2002}. Hence, one may speculate about this value being a universal limit for the recurrence networks of one-dimensional chaotic maps. Moreover, for $x\to 0$ and $x\to 1$, we have $\mathcal{C}_v\to 1$ independent of $\epsilon$. This behaviour is consistent with our previous observations concerning the effect of sharp attractor boundaries on the local clustering coefficient, for example, in case of the tips of the H\'enon attractor (Sec.~\ref{cluster_spatial}).

For the global clustering coefficient, the computation of Eq.~(\ref{cc_global}) leads to
\begin{equation}
\mathcal{C}(\epsilon)=\frac{3}{4}+\epsilon\left(4\ln 2-\frac{5}{2}\right),
\end{equation}
\noindent
which is in good agreement with numerical results (see Figure \ref{ccav_bernoulli}). Hence, deviations from the theoretical value $\frac{3}{4}$ occur exclusively due to boundary effects, and may even lead to $\mathcal{C}=1$ for very large thresholds $\epsilon$. 

\subsection{Logistic map for $a=4$}

The logistic map at $a=4$ is known to have the invariant density
\begin{equation}
p(x)=\frac{1}{\pi}\frac{1}{\sqrt{x(1-x)}}.
\end{equation}
\noindent
With this, we find the following expressions:
\begin{equation}
\begin{split}
&P(A_{v,i}=1,A_{v,j}=1) = \\
&\quad = \left\{
\begin{array}{ll}
\frac{1}{2}-\frac{1}{\pi}\arcsin(1-2x_v-2\epsilon), & 0\leq x_v\leq\epsilon \\
\frac{1}{\pi}\left\{ \arcsin(1-2x_v+2\epsilon)-\arcsin(1-2x_v-2\epsilon) \right\}, & \epsilon\leq x_v\leq 1-\epsilon\\
\frac{1}{2}+\frac{1}{\pi}\arcsin(1-2x_v+2\epsilon), & 1-\epsilon\leq x_v\leq 1
\end{array}
\right.
\end{split}
\end{equation}
\noindent
and
\begin{align}
I_2^{(1)} &= \frac{1}{4}-\frac{1}{2\pi}\arcsin(1-2x_v)-\frac{1}{\pi^2}\int_0^{x_v} dx\ \frac{\arcsin(1-2x-2\epsilon)}{\sqrt{x(1-x)}}, \\
I_2^{(2)} &= \left( \frac{1}{2\pi}-\frac{1}{\pi^2}\arcsin(1-2x_v-2\epsilon) \right) \left( \arcsin(1-2x_v)-\arcsin(1-2\epsilon) \right), \\
I_2^{(3)} &= \frac{1}{\pi^2}\arcsin(1-2x_v-2\epsilon) \left( \arcsin(1-2x_v-2\epsilon)-\arcsin(1-2\epsilon) \right) \nonumber \\
&\qquad +\frac{1}{\pi^2}\int_{\epsilon}^{x_v+\epsilon}dx\ \frac{\arcsin(1-2x+2\epsilon)}{\sqrt{x(1-x)}}, \\
I_2^{(4)} &= \frac{1}{\pi^2}\arcsin(1-2x_v+2\epsilon) \left( \arcsin(1-2x_v+2\epsilon)-\arcsin(1-2x_v) \right) \nonumber \\
&\qquad -\frac{1}{\pi^2}\int_{x_v-\epsilon}^{x_v}dx\ \frac{\arcsin(1-2x-2\epsilon)}{\sqrt{x(1-x)}}, \\
I_2^{(5)} &= \frac{1}{\pi^2}\arcsin(1-2x_v-2\epsilon) \left( \arcsin(1-2x_v-2\epsilon)-\arcsin(1-2x_v) \right) \nonumber \\
&\qquad +\frac{1}{\pi^2}\int_{x_v}^{x_v+\epsilon}dx\ \frac{\arcsin(1-2x+2\epsilon)}{\sqrt{x(1-x)}}, \\
I_2^{(6)} &= \frac{1}{\pi^2}\arcsin(1-2x_v+2\epsilon) \left( \arcsin(1-2x_v+2\epsilon)+\arcsin(1-2\epsilon) \right) \nonumber \\
&\qquad -\frac{1}{\pi^2}\int_{x_v-\epsilon}^{1-\epsilon} \frac{\arcsin(1-2x-2\epsilon)}{\sqrt{x(1-x)}}, \\
I_2^{(7)} &= -\left( \frac{1}{2\pi}+\frac{1}{\pi^2}\arcsin(1-2x_v+2\epsilon) \right) \left( \arcsin(1-2x_v)+\arcsin(1-2\epsilon) \right), \\
I_2^{(8)} &= \frac{1}{4}+\frac{1}{2\pi}\arcsin(1-2x_v)+\frac{1}{\pi^2}\int_{x_v}^1 dx\ \frac{\arcsin(1-2x+2\epsilon)}{\sqrt{x(1-x)}}.
\end{align}
\noindent
Note that the integrals $(I_2^{(1)},I_2^{(2)},I_2^{(3)})$ and $(I_2^{(8)},I_2^{(7)},I_2^{(6)})$ can be transformed into each other by the transformation $x\mapsto 1-x$, and that $I_2^{(4)}+I_2^{(5)}$ is invariant under the same transformation, reflecting the corresponding symmetry of the phase space density $p(x)$.

\begin{figure*}
  \centering
  \includegraphics[scale=0.5]{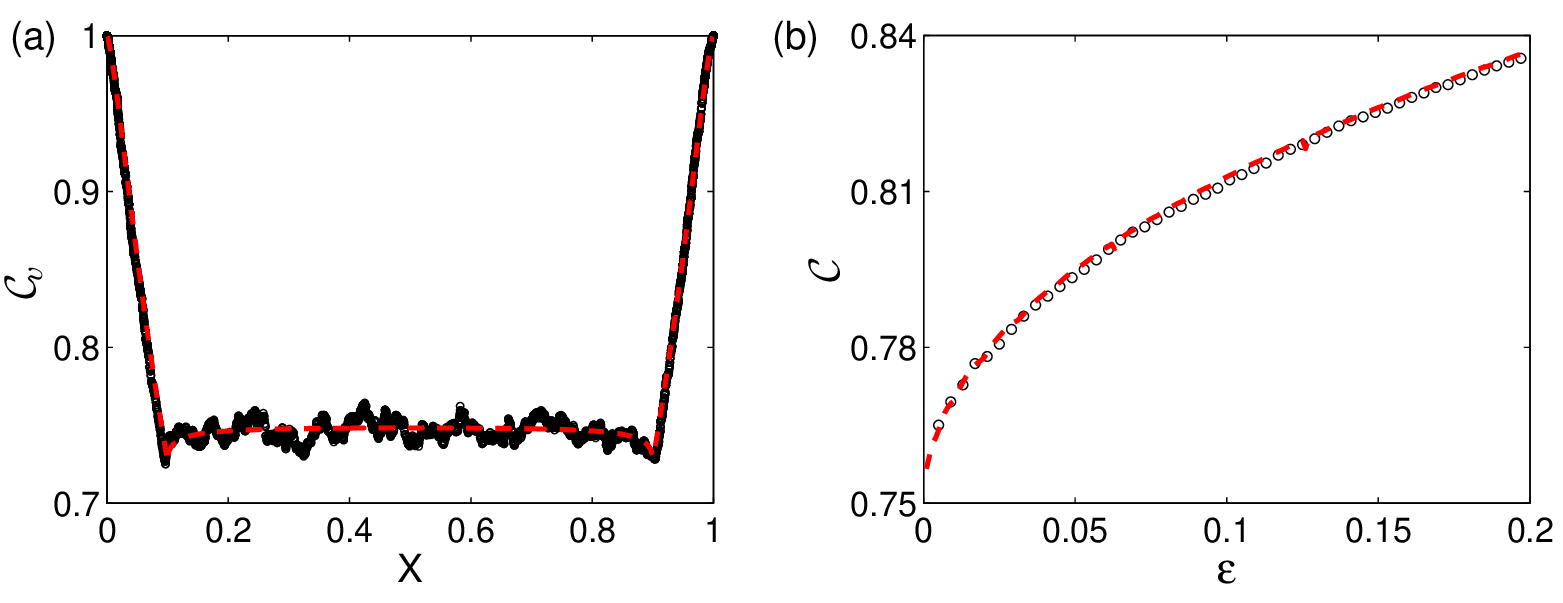}
\caption{\small (a) Analytical (dashed red line) and simulation results (circles) on the dependence of the local clustering coefficient $\mathcal{C}_v$ on the spatial position $x$ for the logistic map at $a=4$ ($N=5,000$, $\epsilon=0.1$). (b) As in (a) for the dependence of the global clustering coefficient $\mathcal{C}$ on $\epsilon$.} \label{cc_logmap}
\end{figure*}

Since the remaining integrals can only be solved numerically, we are not able to give an explicit equation for the functional dependence of the clustering coefficient on both $x$ and $\epsilon$. However, comparing the numerical solution of our analytical results with the clustering coefficients of recurrence networks of one realisation of the logistic map at $a=4$, we find (apart from remaining fluctuations due to the finite length of the considered time series) an excellent agreement (see Fig.~\ref{cc_logmap}). In particular, for the attractor boundaries at $x=0$ and $x=1$, we have $\mathcal{C}_v=1$ independent of $\epsilon$ as for the Bernoulli map. Moreover, we observe almost uniform values of the local clustering coefficient close to $\frac{3}{4}$ within the interval $[\epsilon,1-\epsilon]$, and a systematic tendency towards larger values close to the boundaries. Since the measure of the latter intervals systematically increases with increasing $\epsilon$, these boundary effects are again responsible for the systematic increase in the global clustering coefficient $\mathcal{C}$ as $\epsilon$ becomes larger.

\vspace{1cm}


\end{document}